\documentclass[a4paper,11pt]{article}
\usepackage{jheppub} 
\usepackage{dcolumn}
\usepackage{bm}
\usepackage{amsmath}
\usepackage{amsfonts}
\usepackage{amssymb}
\usepackage{graphicx,subfigure}
\usepackage{color}
\usepackage{latexsym}
\usepackage{slashed} 
\usepackage{mathrsfs}
\usepackage{multirow}
\usepackage{epsfig,psfrag}
\usepackage[scientific-notation=true]{siunitx} 
\usepackage[utf8]{inputenc} 

\graphicspath{{fig/}} 

\makeatletter
\newcommand\figcaption{\def\@captype{figure}\caption} \newcommand\tabcaption{\def\@captype{table}\caption}
\makeatother

\allowdisplaybreaks 

\begin{document}
\title{CEPC Precision of Electroweak Oblique Parameters and Weakly Interacting Dark Matter: the Scalar Case}
\author[a]{Chengfeng Cai,}
\author[b]{Zhao-Huan Yu,}
\author[a]{and Hong-Hao Zhang\footnote{Corresponding author}}

\affiliation[a]{School of Physics, Sun Yat-Sen University, Guangzhou 510275, China}
\affiliation[b]{ARC Centre of Excellence for Particle Physics at the Terascale,
School of Physics, The University of Melbourne, Victoria 3010, Australia}

\emailAdd{caichf@mail2.sysu.edu.cn}
\emailAdd{zhao-huan.yu@unimelb.edu.au}
\emailAdd{zhh98@mail.sysu.edu.cn}

\abstract{We investigate the sensitivity to weakly interacting scalar dark matter from future determination of electroweak oblique parameters in the Circular Electron-Positron Collider (CEPC) project.
As illuminating examples, three dark matter models with scalar electroweak multiplets are studied.
The multiplet couplings to the standard model Higgs doublet can break the mass degeneracy among the components, leading to nonzero contributions to oblique parameters.
The dark matter candidate in these model is either a CP-even or CP-odd scalar, whose trilinear coupling to the Higgs boson could induce direct detection signals.
For some moderate coupling values, we find that the CEPC sensitivity can be better than current direct detection experiments, exploring up to a TeV mass scale.}

\maketitle

\section{Introduction}

The discovery of the Higgs boson~\cite{Aad:2012tfa,Chatrchyan:2012xdj} at the Large Hadron Collider (LHC) confirms the mechanism of electroweak (EW) symmetry breaking in the standard model (SM). However, the SM is far away from a complete description of particle physics. It fails to explain neutrino masses and dark matter (DM). Moreover, the radiative correction to the Higgs boson mass gives rise to the gauge hierarchy problem, implying a new physics scale not far above the weak scale.
As the latest discovered fundamental particle, the Higgs boson is also the least known one. An urgent task after the discovery is to study its precise interaction properties, which are essential for understanding new physics that provides solutions to the problems in the SM.

Because of the enormous backgrounds in hadron collisions, the LHC is incapable of pinning down the Higgs couplings with desired precisions. Therefore, in order to accurately measure most of the Higgs couplings, we needs an $e^+e^-$ collider serving as a Higgs factory, which should have high luminosity and operate at $\sqrt{s}\sim 240~\si{GeV}$ to maximize the $e^+e^-\to Zh$ production rate.
The Circular Electron-Positron Collider (CEPC)~\cite{CEPC-SPPCStudyGroup:2015csa}, recently proposed by the Chinese high energy physics community, fulfills these requirements. In addition, there are also plans for low energy CEPC runs around the $Z$ pole and around the $WW$ threshold. It is expected to produce up to $10^{10}$ $Z$ bosons, which is a thousand times of the $Z$ boson events observed at the LEP.
Therefore, CEPC would greatly improve the measurements of EW precision observables since the end of LEP runs. Any deviation from the SM in such ultra-high precision tests would point to new physics.
Recent estimations of the CEPC sensitivity to new physics via precision tests can be found in Refs.~\cite{McCullough:2013rea,Hu:2014eia,Fan:2014vta,Fan:2014axa,Harigaya:2015yaa,Shen:2015pha,Cao:2014ita,Cao:2015iua,Fedderke:2015txa,Gori:2015nqa,Huang:2015izx,Ge:2016zro,Cao:2016qgc,Kobakhidze:2016mfx,Cai:2016sjz,Cao:2016uwt,Wu:2017kgr}.

Although what dark matter is and how it interacts are open questions.
There are many reasons for believing that DM consists of new particles beyond the SM~\cite{Jungman:1995df,Bertone:2004pz,Feng:2010gw}.
The most popular class of DM candidates is weakly interacting massive particles (WIMPs).
WIMPs are assumed to interact with SM particles through the weak interaction or through some unknown interactions whose strength is similar to the weak interaction.
As a result, thermal production of WIMPs in the early Universe provides a relic abundance in agreement with observation.
If no new gauge interaction is introduced, a reasonable choice for WIMP model building is to consider a dark sector containing EW $\mathrm{SU}(2)_\mathrm{L}$ multiplets.
The lightest mass eigenstate of the neutral components in such multiplets can be stabilized by a $Z_2$ symmetry, becoming an appealing DM candidate.

Such multiplets would induce loop effects on EW precision tests at the CEPC. In particular, they could contribute to EW oblique parameters $S$, $T$, and $U$~\cite{Peskin:1990zt,Peskin:1991sw}, which quantify the effects from new particles that couple to EW gauge bosons but do not directly couple to SM fermions. Since the $Z_2$ symmetry typically forbids the couplings of dark sector multiplets to SM fermions, the oblique parameters could provide a reasonable test on this kind of DM models.

In a previous work~\cite{Cai:2016sjz}, we estimated the CEPC precision of EW oblique parameters and investigate the corresponding sensitivity to DM models with EW fermionic multiplets. In this work, we extend the study to EW scalar multiplets.
Since the oblique parameters are dimensionless measures of EW symmetry breaking, only the $\mathrm{SU}(2)_\mathrm{L}$ multiplets whose components are split in mass can induce nonzero contributions~\cite{Zhang:2006de,Zhang:2006vt}.
In the fermionic case, Yukawa couplings to the SM Higgs doublet are needed to break the mass degeneracy after the Higgs field acquires a vacuum expectation value (VEV).
Therefore, in the previous study~\cite{Cai:2016sjz} we discussed two vector-like dark sector multiplets living in two $\mathrm{SU}(2)_\mathrm{L}$ representations whose dimensions differ by one for allowing such Yukawa couplings~\cite{Cohen:2011ec,Dedes:2014hga,Tait:2016qbg}.

On the other hand, a scalar multiplet $\Phi$ can have quartic couplings to the Higgs doublet $H$.
But it should be noted that a real scalar multiplet has only one such quartic coupling, which is in the form of $\lambda'\Phi^\dag\Phi H^\dag H$ and just induces a common mass shift to every components.
Thus, the mass degeneracy cannot be broken if the dark sector only contains a real scalar multiplet (see, \textit{e.g.}, Refs.~\cite{Cirelli:2005uq,Hambye:2009pw,Cai:2015kpa}).
Nevertheless, in general this does not apply to a complex scalar multiplet (see, \textit{e.g.}, Refs.~\cite{Cirelli:2005uq,Hambye:2009pw,Cai:2012kt,Earl:2013jsa,AbdusSalam:2013eya,Logan:2016ivc,Chowdhury:2016mtl}), because it can have couplings like $\lambda''\Phi^\dag\tau^a\Phi H^\dag \sigma^a H$ ($\tau^a$ are the $SU(2)$ generators for $\Phi$ and $\sigma$ is the Pauli matrices), which can split its components after EW symmetry breaking.
If there are more than one scalar multiplets, the interactions with the Higgs doublet will be more complicated, leading to extra trilinear and quartic couplings that could break the mass degeneracy.
Some studies on DM models with two EW scalar multiplets can be found in Refs.~\cite{Cohen:2011ec,Fischer:2013hwa,Cheung:2013dua,Kakizaki:2016dza,Lu:2016dbc}.

As illuminating examples, we study the following WIMP models with $\mathrm{SU}(2)_\mathrm{L}$ scalar multiplets:
\begin{itemize}
\item Singlet-Doublet Scalar Dark Matter (SDSDM): a real singlet scalar  with zero hypercharge ($Y=0$) and a complex doublet scalar with $Y=1/2$~\cite{Cohen:2011ec,Cheung:2013dua,Lu:2016dbc};
\item Singlet-Triplet Scalar Dark Matter (STSDM): a real singlet scalar with $Y=0$ and a complex triplet scalar with $Y=0$~\cite{Lu:2016dbc};
\item Quadruplet Scalar Dark Matter (QSDM): a complex quadruplet scalar field with $Y=1/2$~\cite{Cirelli:2005uq,AbdusSalam:2013eya,Chowdhury:2016mtl}.
\end{itemize}
To stabilize the DM particle, these dark sector scalar fields are set to be odd under a $Z_2$ symmetry, while the SM fields are even.
The lightest mass eigenstate of the neutral components, which can be CP-even or CP-odd, plays the role of the DM candidate.
For making this candidate stable, it also requires that the charged particles in the dark sector should be heavier than this particle.

The organization of this paper is as follows. In Section~\ref{sec:STU} we discuss EW oblique parameters and their CEPC precision. In Sections~\ref{sec:SDSDM}, \ref{sec:STSDM}, and \ref{sec:QSDM}, we investigate the expected sensitivity on the SDSDM, STSDM, and QSDM models from the oblique parameters, respectively. For comparison, current constraints and future sensitivity from DM direct detection experiments are also demonstrated. The conclusion is given in Section~\ref{sec:concl}.
Appendix~\ref{app:conv_tensor} describes our convention for $\mathrm{SU}(2)$ tensors, while Appendix~\ref{app:DM_scat} provides supplementary formulas for DM-nucleon scatterings.

\section{Electroweak Oblique Parameters and CEPC Precision}
\label{sec:STU}

EW oblique parameters $S$, $T$, and $U$ describe loop corrections to EW gauge boson propagators. Their definitions are given by~\cite{Peskin:1990zt,Peskin:1991sw}
\begin{equation}
S=16\pi[\Pi'_{33}(0)-\Pi'_{3Q}(0)],~~
T=\frac{4\pi}{s_\mathrm{W}^2c_\mathrm{W}^2m_Z^2}[\Pi_{11}(0)-\Pi_{33}(0)],~~
U=16\pi[\Pi'_{11}(0)-\Pi'_{33}(0)],
\end{equation}
where the derivatives are defined as $\Pi'_{IJ}(0)\equiv\partial\Pi_{IJ}(p^2)/\partial p^2|_{p^2=0}$.
$\Pi_{3Q}(p^2)$, $\Pi_{11}(p^2)$, and $\Pi_{33}(p^2)$ are the $g_{\mu\nu}$ coefficients of the vacuum polarization amplitudes in the gauge basis induced by new physics.
Their relations to the vacuum polarization amplitudes in the physical basis are
\begin{eqnarray}
\Pi_{WW}(p^2)&=&\frac{e^2}{s_\mathrm{W}^2}\Pi_{11}(p^2),\quad
\Pi_{AA}(p^2)=e^2\Pi_{QQ}(p^2),\\
\Pi_{ZZ}(p^2)&=&\frac{e^2}{s_\mathrm{W}^2c_\mathrm{W}^2}[\Pi_{33}(p^2)-2s_\mathrm{W}^2\Pi_{3Q}(p^2)+s_\mathrm{W}^4\Pi_{QQ}(p^2)],\\
\Pi_{ZA}(p^2)&=&\frac{e^2}{s_\mathrm{W}c_\mathrm{W}}[\Pi_{3Q}(p^2)-s_\mathrm{W}^2\Pi_{QQ}(p^2)].
\end{eqnarray}
As the usual convention, the shortcuts $s_\mathrm{W}\equiv\sin\theta_\mathrm{W}$ and $c_\mathrm{W}\equiv\cos\theta_\mathrm{W}$, where $\theta_\mathrm{W}$ is the Weinberg angle.

By these definitions, $S$, $T$, and $U$ are dimensionless.
As we only care about new physics effects, the standard model corresponds to $S=T=U=0$.
At the lowest order of effective field theory, $S$ and $T$ are corresponding to dimension-6 operators ${H^\dag }W_{\mu \nu }^a{\sigma ^a}H{B^{\mu \nu }}$ and ${H^\dag }({D_\mu }H)({D^\mu }H )^\dag H$, respectively, while $U$ is corresponding to a dimension-8 operator ${H^\dag }W_{\mu \nu }^a{\sigma ^a}H{H^\dag }{W^{b\mu \nu }}{\sigma ^b}H$~\cite{Han:2008es}.
Therefore, $U$ is expected to be much smaller than the other two in many new physics models.
Global EW fits are often carried out under the assumption $U=0$.

New $\mathrm{SU}(2)_\mathrm{L}$ multiplets would generally contribute to the oblique parameters.
However, these contributions may vanish under specific situations.
First, the oblique parameters respond to EW symmetry breaking, in particular, to the mass splittings among the multiplet components induced by the nonzero Higgs VEV. If the VEV just gives a common mass shift to every components in a multiplet, the effect can be absorbed into the gauge-invariant mass term of the multiplet. Consequently, no EW symmetry breaking effect manifests, leading to vanishing $S$, $T$, and $U$.
Second, by definition $S$ relates to the $\mathrm{U}(1)_\mathrm{Y}$ gauge field. As a result, a scalar multiplet or a vector-like fermionic multiplet with zero hypercharge cannot contribute to $S$~\cite{Zhang:2006de,Zhang:2006vt}.
Third, if the interactions between the Higgs field and multiplets respect a custodial global symmetry, $T$ and $U$ will vanish.

Let us take a closer look at the third situation. In this case, before EW symmetry breaking there is an $\mathrm{SU}(2)_\mathrm{L}\times\mathrm{SU}(2)_\mathrm{R}$ global symmetry in the SM Higgs potential terms and the interaction terms between the Higgs doublet $H$ and new $\mathrm{SU}(2)_\mathrm{L}$ multiplets.
If one expresses the Higgs field in an $\mathrm{SU}(2)_\mathrm{L}\times\mathrm{SU}(2)_\mathrm{R}$ bidoublet form $\mathcal{H} = (\tilde{H},H)$ with $\tilde{H}\equiv i\sigma^2 H^*$, the symmetry means that the related terms are invariant under the global transformation $\mathcal{H}\to U_\mathrm{L}\mathcal{H}U_\mathrm{R}^\dag$ where $U_\mathrm{L}\in \mathrm{SU}(2)_\mathrm{L}$ and $U_\mathrm{R}\in \mathrm{SU}(2)_\mathrm{R}$.
After EW symmetry breaking, the Higgs field get a VEV $v$ and $\left<\mathcal{H}\right>=\mathrm{diag}(v,v)/\sqrt{2}$. Now if $U_\mathrm{L}=U_\mathrm{R}$, we will have $\left<\mathcal{H}\right>\to\left<\mathcal{H}\right>$ and the vacuum will be invariant. Thus, the $\mathrm{SU}(2)_\mathrm{L}\times\mathrm{SU}(2)_\mathrm{R}$ symmetry is broken to an $\mathrm{SU}(2)_\mathrm{L+R}$ symmetry with equal $\mathrm{SU}(2)_\mathrm{L}$ and $\mathrm{SU}(2)_\mathrm{R}$ rotations.
The $\mathrm{SU}(2)_\mathrm{L}$ gauge bosons $W^a_\mu$ ($a=1,2,3$) transform as a triplet under $\mathrm{SU}(2)_\mathrm{L+R}$.
Consequently, this symmetry ensures that $W^1_\mu$, $W^2_\mu$, and $W^3_\mu$ receive identical corrections in vacuum polarizations from the new multiplets, resulting in $\Pi_{11}(p^2) = \Pi_{33}(p^2)$ and thus $T=U=0$.
Such an $\mathrm{SU}(2)_\mathrm{L+R}$ global symmetry is called a ``custodial'' symmetry because it protects the tree-level relation $\rho\equiv m_W^2/(m_Z^2 c_\mathrm{W}^2)=1$
up to radiative corrections~\cite{Sikivie:1980hm}.
This conclusion can be easily understood through the loop-level relation $\rho-1=\alpha T$~\cite{Peskin:1991sw,Zhang:2009rm}.

Current constraints on the EW oblique parameters from global fits and estimations of future precisions can be found in Refs.~\cite{Ciuchini:2013pca,Baak:2014ora,Fan:2014vta,CEPC-SPPCStudyGroup:2015csa,deBlas:2016ojx,Cai:2016sjz}.
Below we adopt the results for the CEPC precision estimated in our previous work~\cite{Cai:2016sjz}.
If all of $S$, $T$, and $U$ are free parameters in the global fits,
we have the following results.
\begin{itemize}
\item Current precision: \\*
\hspace*{.5em} $\sigma_S = 0.10$,~ $\sigma_T = 0.12$,~ $\sigma_U = 0.094$,~ $\rho_{ST} =  + 0.89$,~ $\rho_{SU} =  - 0.55$,~ $\rho_{TU} =  - 0.80$;
\item CEPC-B precision: \\*
\hspace*{.5em} $\sigma_S = 0.021$,~ $\sigma_T = 0.026$,~ $\sigma_U = 0.020$,~ $\rho_{ST} =  + 0.90$,~ $\rho_{SU} =  - 0.68$,~ $\rho_{TU} =  - 0.84$;
\item CEPC-I precision: \\*
\hspace*{.5em} $\sigma_S = 0.011$,~ $\sigma_T = 0.0071$,~ $\sigma_U = 0.010$,~ $\rho _{ST} =  + 0.74$,~ $\rho_{SU} =  + 0.15$,~ $\rho_{TU} =  - 0.21$.
\end{itemize}
Here $\sigma_i$ and $\rho_{ij}$ ($i,j=S,T,U$) denote standard deviations and correlation coefficients for the oblique parameters, respectively.

The ``current'' precision was derived from the latest experimental data for EW precision observables.
The ``CEPC-B'' precision stands for the baseline precision of the CEPC, which was estimated taking into account the experimental uncertainties that could be reduced by CEPC runs, as well as the reduced theoretical uncertainties via fully calculating 3-loop corrections in the future.
The ``CEPC-I'' precision was evaluated assuming two additional improvements.
One is a high-precision beam energy calibration for improving the measurements of the $Z$ boson mass and decay width at the CEPC.
The other one requires a $t\bar{t}$ threshold scan provided by other $e^+e^-$ colliders, such as ILC and FCC-ee.

We also list the following fit results obtained by fixing some of the oblique parameter to zero, which could be more suitable for particular models.
\begin{enumerate}
\item $U=0$ fixed
\begin{itemize}
\item Current precision: $\sigma_S = 0.085$,~ $\sigma_T = 0.072$,~ $\rho_{ST} =  + 0.90$;
\item CEPC-B precision: $\sigma_S = 0.015$,~ $\sigma_T = 0.014$,~ $\rho_{ST} =  + 0.83$;
\item CEPC-I precision: $\sigma_S = 0.011$,~ $\sigma_T = 0.0069$,~  $\rho_{ST} =  + 0.80$.
\end{itemize}
\item $S=0$ fixed
\begin{itemize}
\item Current precision: $\sigma_T = 0.054$,~ $\sigma_U = 0.078$,~ $\rho_{TU} = -0.81$;
\item CEPC-B precision: $\sigma_T = 0.011$,~ $\sigma_U = 0.015$,~ $\rho_{TU} = -0.72$;
\item CEPC-I precision: $\sigma_T = 0.0048$,~ $\sigma_U = 0.010$,~  $\rho_{TU} = -0.48$.
\end{itemize}
\item $S=U=0$ fixed
\begin{itemize}
\item Current precision: $\sigma_T = 0.032$;
\item CEPC-B precision: $\sigma_T = 0.0079$;
\item CEPC-I precision: $\sigma_T = 0.0042$.
\end{itemize}
\end{enumerate}
As mentioned above, the results with $U=0$ fixed is useful for many new physics models predicting $U\ll T,S$.
Moreover, since both the singlet and triplet scalars in the STSDM model, which will be discussed in Section~\ref{sec:STSDM}, have zero hypercharge and would not contribute to $S$, we will use the fit results with $S=0$ fixed and with $S=U=0$ fixed for this model.
Below we discuss the sensitivity to scalar WIMP DM models based on these global fit results.

\section{Singlet-Doublet Scalar Dark Matter}
\label{sec:SDSDM}

\subsection{Fields and Interactions}

In the SDSDM model, we introduce a real singlet scalar $S$ and a complex doublet scalar $\Phi$ with the denoted $(\mathrm{SU}(2)_\mathrm{L},\mathrm{U}(1)_Y)$ gauge transformation properties:
\begin{equation}
S\in (\mathbf{1},0), \quad
\Phi=\begin{pmatrix}\phi^+\\
(\phi^0+ia^0)/\sqrt{2}\end{pmatrix}\in(\mathbf{2},1/2).
\end{equation}
Both are odd under a $Z_2$ symmetry for ensuring the stability  of the DM particle.
Their kinetic and interacting properties are encoded in the Lagrangian:
\begin{eqnarray}
\mathcal{L}&\supset&
\frac{1}{2}(\partial_\mu S)^2-\frac{1}{2}m_S^2S^2
+(D_\mu\Phi)^\dag D^\mu\Phi-m_D^2|\Phi|^2
-(\kappa S\Phi^\dag H+\mathrm{h.c.})
-\frac{\lambda_{Sh}}{2}S^2|H|^2
\nonumber\\
&&-\lambda_1|H|^2|\Phi|^2
-[\lambda_2(\Phi^\dag H)^2+\mathrm{h.c.}]-\lambda_3|\Phi^\dag H|^2+ (\text{irrelevant terms}),\qquad
\end{eqnarray}
where $H$ is the SM Higgs doublet and the covariant derivative $D_\mu = \partial_\mu -i g W_\mu^a\tau^a-ig'B_\mu/2$ with $\tau^a=\sigma^a/2$.
Note that $\kappa$ has a mass dimension.
We only present the terms corresponding to masses, gauge interactions, and couplings to the Higgs field for $S$ and $\Phi$.
The self interaction terms and the couplings only between $S$ and $\Phi$ are not relevant throughout the analysis.
The gauge interaction terms for $\Phi$ can be found in Eq.~\eqref{eq:SD:gauge}.

In the unitary gauge, we have $H=(0,(v+h)/\sqrt{2})^\mathrm{T}$, leading to the following mass terms:
\begin{equation}
\mathcal{L}_\mathrm{mass}=-\frac{1}{2}(S\quad\phi^0)M_0^2\begin{pmatrix}S\\ \phi^0\end{pmatrix}
-\frac{1}{2}m_a^2(a^0)^2
-m_C^2|\phi^+|^2,
\end{equation}
with
\begin{eqnarray}
&&M_0^2=\begin{pmatrix}m_S^2+\dfrac{1}{2}\lambda_{Sh}v^2&\kappa v\\[.4em] \kappa v&m_D^2+\dfrac{1}{2}(\lambda_1+2\lambda_2+\lambda_3)v^2\end{pmatrix},\\
&&m_a^2=m_D^2+\frac{1}{2}(\lambda_1-2\lambda_2+\lambda_3)v^2,\quad
m_C^2=m_D^2+\frac{1}{2}\lambda_1v^2.
\end{eqnarray}
Two physical CP-even neutral scalars $X_1$ and $X_2$ are obtained by diagonalizing the matrix $M_0^2$ through
\begin{equation}
U^\mathrm{T}M_0^2U=\begin{pmatrix}m_1^2&0\\0&m_2^2\end{pmatrix},\quad
U=\begin{pmatrix}c_\theta&-s_\theta\\s_\theta&c_\theta\end{pmatrix},\quad
\begin{pmatrix}S\\ \phi^0\end{pmatrix}=U\begin{pmatrix}X_1\\ X_2\end{pmatrix},
\end{equation}
where $c_\theta\equiv\cos\theta$, $s_\theta\equiv\sin\theta$, and the rotation angle $\theta$ satisfies
\begin{equation}
\tan 2\theta = \frac{2 (M_0^2)_{12}}{(M_0^2)_{22}-(M_0^2)_{11}}.
\end{equation}
The masses of $X_1$ and $X_2$ are
\begin{equation}
m_{1,2}^2=\frac{1}{2}\Big\{(M_0^2)_{11}+(M_0^2)_{22}\mp\sqrt{[(M_0^2)_{11}-(M_0^2)_{22}]^2+4(M_0^2)_{12}^2}\Big\}.
\end{equation}

Since the DM candidate should be a neutral particle that is the lightest particle in the dark sector, there are three different situations.
\begin{itemize}
\item $m_1<m_a,m_C$: the DM candidate is the lighter CP-even scalar $X_1$;
\item $m_a<m_1,m_C$: the DM candidate is the CP-odd particle $a^0$;
\item $m_C<m_1,m_a$: there is no stable DM candidate.
\end{itemize}
The $hX_1X_1$ and $ha^0a^0$ trilinear couplings can be read off from the Lagrangian as
\begin{equation}\label{SDtrilinear}
\mathcal{L}\supset -\left[\frac{\kappa}{v}s_\theta c_\theta+\frac{\lambda_{Sh}}{2}c_\theta^2+\frac{1}{2}(\lambda_1+2\lambda_{2}+\lambda_{3})s_\theta^2\right]vhX_1^2
-\frac{1}{2}(\lambda_1-2\lambda_{2}+\lambda_{3})vh(a^0)^2.
\end{equation}
They could induce spin-independent DM-nucleon scatterings according to Appendix~\ref{app:DM_scat}, and hence could be probed in DM direct detection experiments.

Dark sector scalars affect the vacuum polarizations of EW gauge bosons at one-loop level, and thus may contribute to the EW oblique parameters $S$, $T$, and $U$.
Eqs.~\eqref{eq:SD:Pi_33}--\eqref{eq:SD:Pi_11} give their contributions to $\Pi_{IJ}$, whose numerical values are calculated with \texttt{LoopTools}~\cite{Hahn:1998yk}.

\subsection{Custodial Symmetry}

The custodial symmetry exists when the couplings satisfy certain conditions.
To find out such conditions, we can construct two $\mathrm{SU}(2)_\mathrm{L}\times \mathrm{SU}(2)_\mathrm{R}$ bidoublets from $H$ and $\Phi$:
\begin{equation}
\mathcal{H}=(\tilde{H},H),\quad \mathcal{F}=(\tilde{\Phi},\Phi),
\end{equation}
with $\tilde H\equiv i\sigma^2H^*$ and $\tilde \Phi\equiv i\sigma^2\Phi^*$.
Thus we have the following $\mathrm{SU}(2)_\mathrm{L}\times \mathrm{SU}(2)_\mathrm{R}$ invariants
\begin{equation}
\mathcal{Z}_1\equiv\mathrm{tr}[\mathcal{H}^\dag \mathcal{H}]=2|H|^2,~~
\mathcal{Z}_2\equiv\mathrm{tr}[\mathcal{F}^\dag \mathcal{F}]=2|\Phi|^2,~~
\mathcal{Z}_3\equiv\mathrm{tr}[\mathcal{F}^\dag \mathcal{H}]=\Phi^\dag H+H^\dag\Phi.
\end{equation}
Therefore, if
\begin{equation}
\lambda_3=2\lambda_2,
\end{equation}
the scalar potential will respect the custodial symmetry:
\begin{eqnarray}
V&\supset&(\kappa S\Phi^\dag H+\mathrm{h.c.})
+\frac{\lambda_{Sh}}{2}S^2|H|^2
+\lambda_1|H|^2|\Phi|^2
+[\lambda_2(\Phi^\dag H)^2+\mathrm{h.c.}]
+\lambda_3|\Phi^\dag H|^2\nonumber\\
&=&\kappa S\mathcal{Z}_3
+\frac{\lambda_{Sh}}{4} S^2\mathcal{Z}_1
+\frac{\lambda_1}{4} \mathcal{Z}_1\mathcal{Z}_2
+\lambda_2\mathcal{Z}_3^2.
\end{eqnarray}

By defining $\mathcal{F}=(-\tilde{\Phi},\Phi)$ instead, one can easily examine that $\lambda_3=-2\lambda_2$ and $\kappa=0$ is another condition for custodial symmetry.
Both conditions lead to $T=U=0$.
Besides, if $\lambda_2=\lambda_3=0$ and $\kappa=0$, the mass terms become
\begin{equation}
{\mathcal{L}_{{\mathrm{mass}}}} =  - \frac{1}{2}\left( {m_S^2 + \frac{1}{2}{\lambda _{Sh}}{v^2}\,} \right){S^2} - \left( {m_D^2 + \frac{1}{2}{\lambda _1}{v^2}} \right)\left[ {\frac{1}{2}{{({\phi ^0})}^2} + \frac{1}{2}{{({a^0})}^2} + |{\phi ^ + }{|^2}} \right].
\end{equation}
There will be no mixing between the CP-even neutral scalars, and the components of $\Phi$ have degenerate masses. Thus, the masses induced by $\lambda_{Sh}$ and $\lambda_1$ can be absorbed into $m_S$ and $m_D$, which come from the gauge-invariant mass terms that still respect the $\mathrm{SU}(2)_L\times\mathrm{U}(1)_\mathrm{Y}$ gauge symmetry after the Higgs field develops a nonzero VEV. Consequently, no EW symmetry breaking effect manifests in oblique corrections, resulting in $S=T=U=0$.

\subsection{Expected Sensitivity}

\begin{figure}[!t]
\centering
\includegraphics[width=0.49\textwidth]{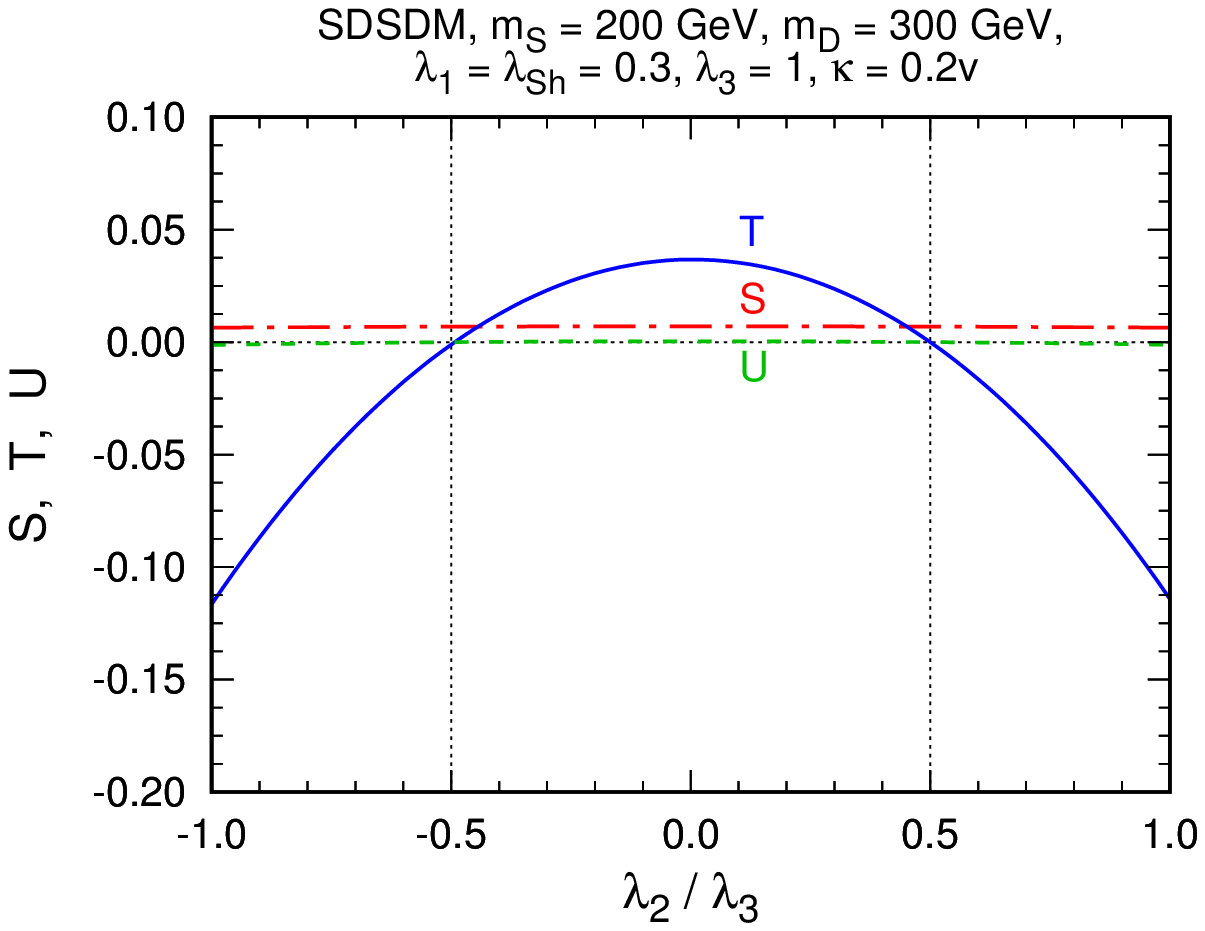}
\includegraphics[width=0.49\textwidth]{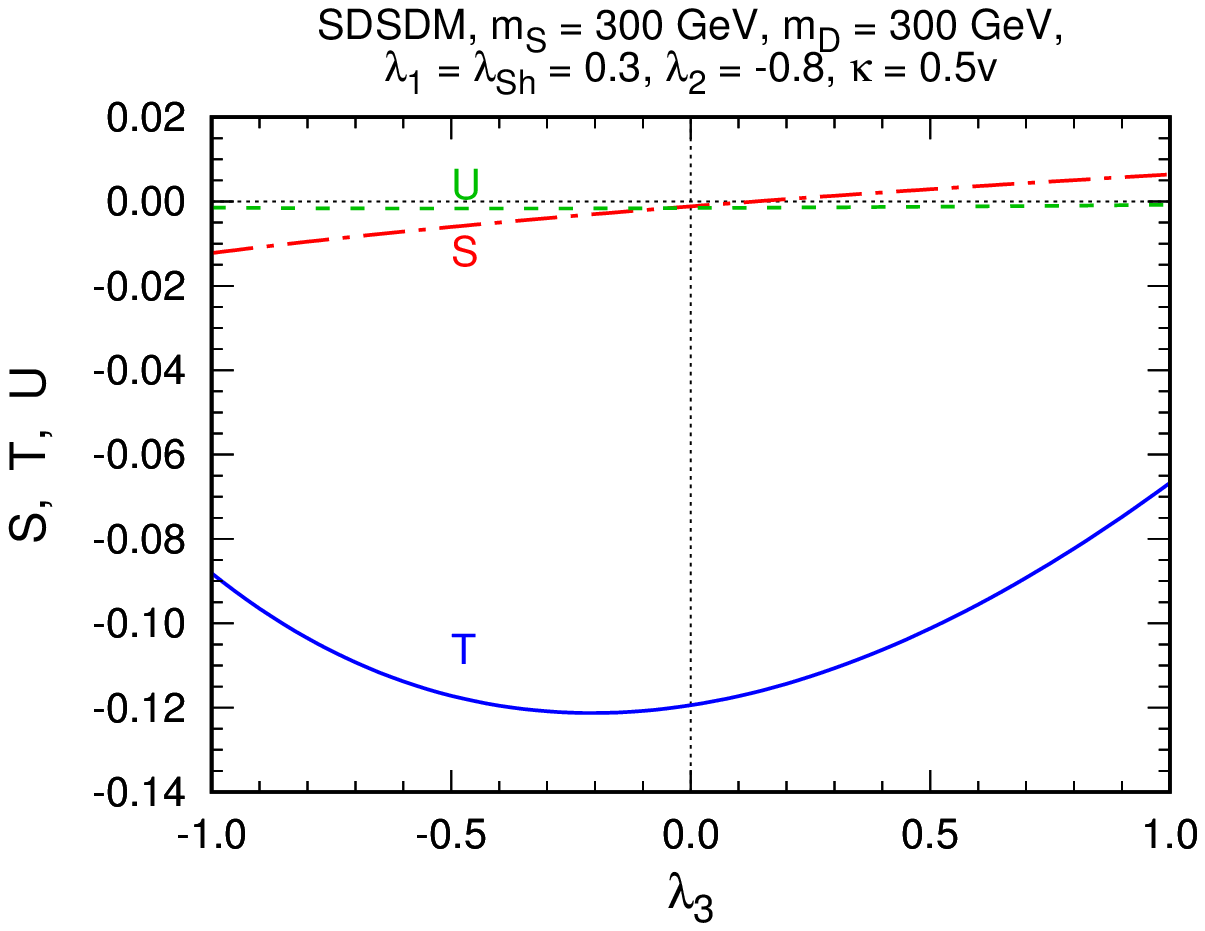}
\caption{$S$, $T$, and $U$ as functions of $\lambda_2/\lambda_3$ (left) and $\lambda_3$ (right) in the SDSDM model. In the left panel, the parameters are fixed as $m_S=200~\si{GeV}$, $m_D=300~\si{GeV}$, $\lambda_1=\lambda_{Sh}=0.3$, $\lambda_3=1$, and $\kappa=0.2v$.
In the right panel, the parameters are fixed as $m_S=m_D=300~\si{GeV}$, $\lambda_1=\lambda_{Sh}=0.3$, $\lambda_2=-0.8$, and $\kappa=0.5v$.}
\label{SDSTU}
\end{figure}

Below we discuss the prediction of EW oblique parameters and the CEPC sensitivity to this model.
In the left panel of Fig.~\ref{SDSTU}, we fix some parameters and show how $S$, $T$, and $U$ vary with the ratio $\lambda_2/\lambda_3$. As expected, $T$ and $U$ vanish at $\lambda_2/\lambda_3=0.5$ due to the custodial symmetry.
They also vanish around $\lambda_2/\lambda_3=-0.5$, where the custodial symmetry is approximately respected since $\kappa$ has been set to a small coupling.
The magnitude of $T$ becomes large once it is away from the custodial symmetric point, while $U$ and $S$ remain small as $\lambda_2/\lambda_3$ varies.

The right panel of the Fig.~\ref{SDSTU} shows the oblique parameters as functions of $\lambda_3$ with the other parameters fixed.
If $\lambda_2=0$ and $\kappa=0$, we should expect $S$ reaches zero at $\lambda_3=0$.
Since here $\lambda_2$ and $\kappa$ are nonzero, $S$ crosses the zero value at a point slightly deviating from $\lambda_3=0$.
Note that $\lambda_3$ is the dominant contribution to the $S$ parameter, and its sign determines the sign of $S$.

\begin{figure}[!t]
\centering
\includegraphics[width=0.49\textwidth]{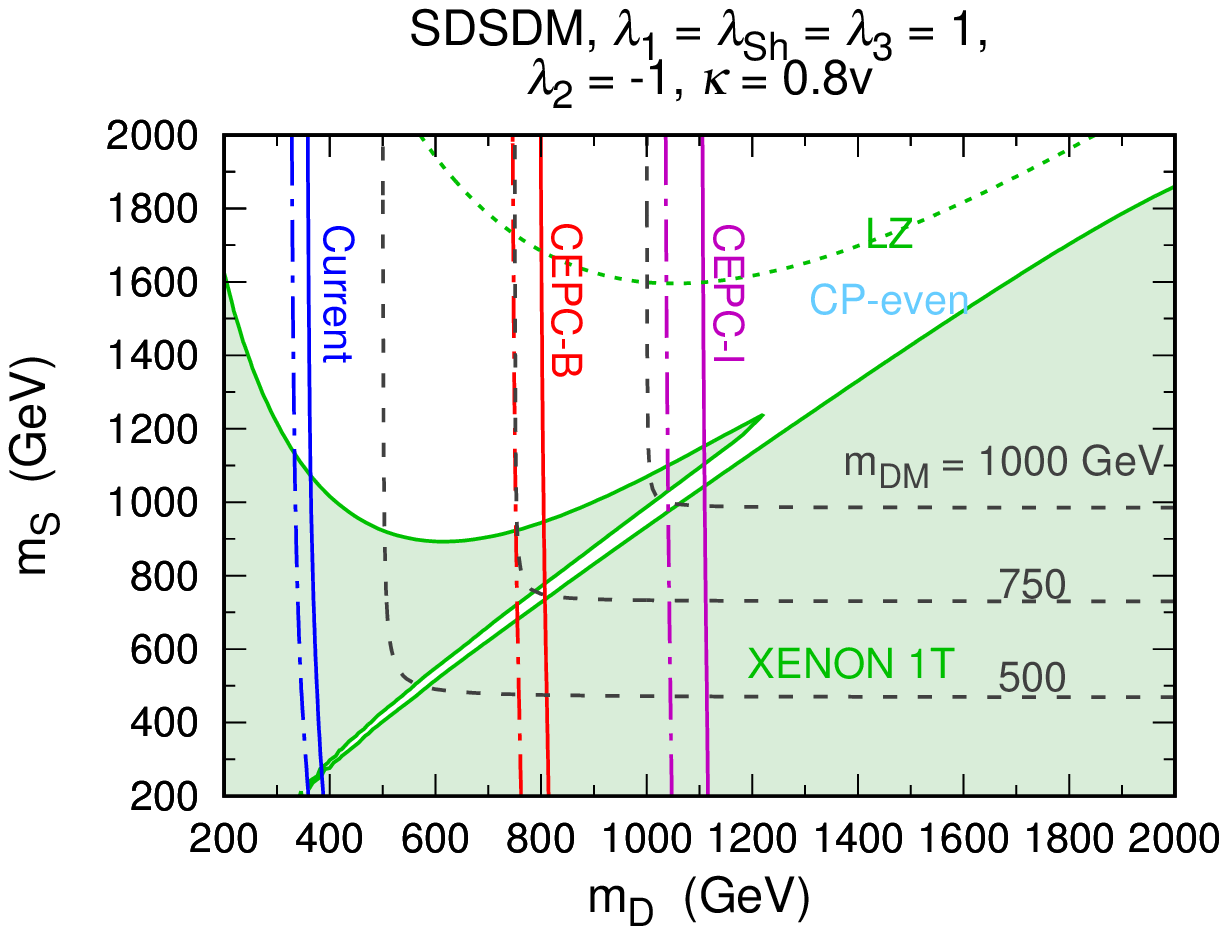}
\includegraphics[width=0.49\textwidth]{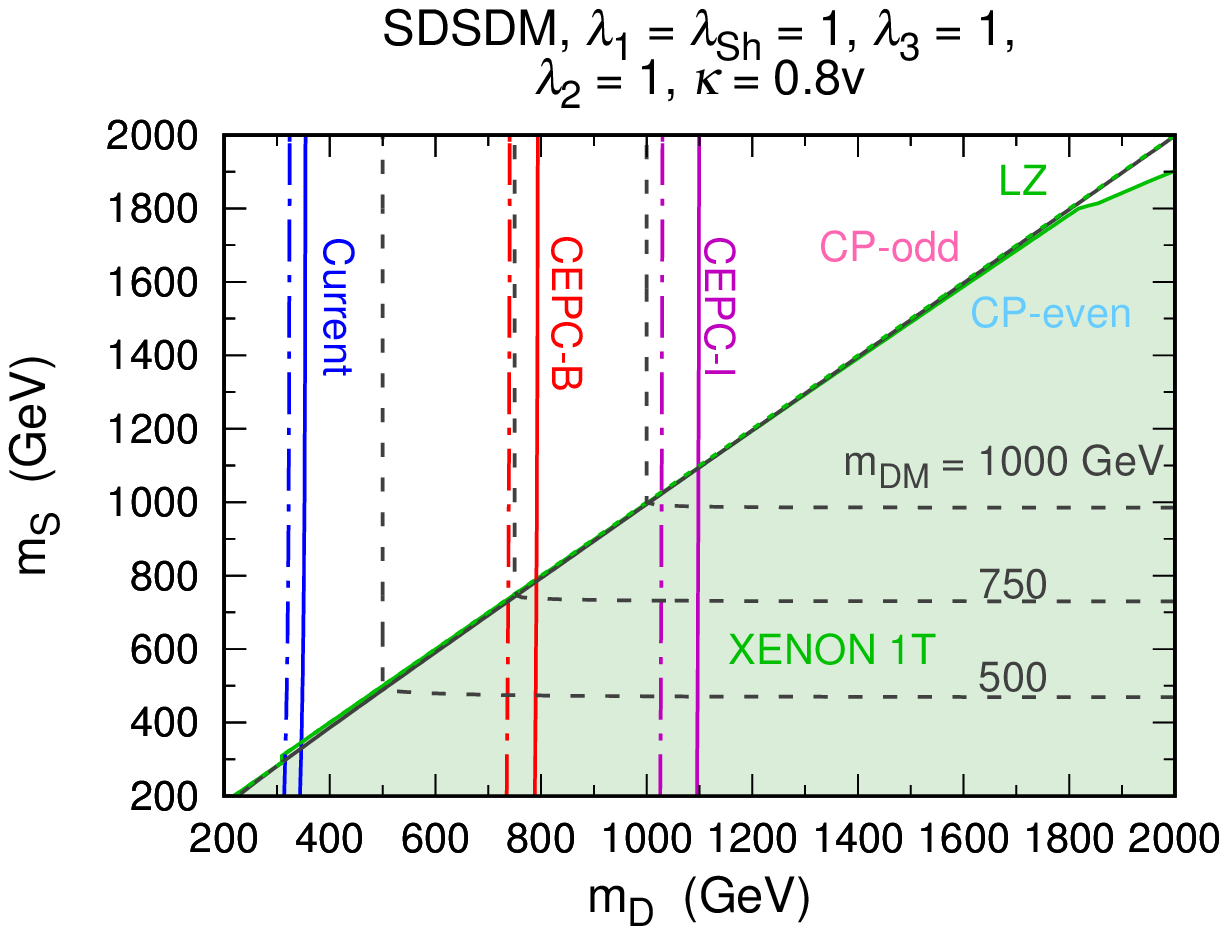}
\caption{Expected 95\% CL sensitivities from CEPC precisions of EW oblique parameters in the $m_D-m_S$ plane for the SDSDM model.
In the left (right) panel, the couplings are fixed as $\lambda_1=\lambda_{Sh}=\lambda_3=1$, $\lambda_2=-1$ ($\lambda_2=1$), and $\kappa=0.8v$.
The 95\% CL sensitivities from current, CEPC-I, and CEPC-B precisions are indicated by blue, red, and purple colors, respectively, while solid (dot-dashed) lines corresponds to the global EW fits with $U=0$ (with free oblique parameters). Filled green regions are excluded at 90\% CL by the XENON1T direct detection experiment~\cite{Aprile:2017iyp}, while green dotted lines denote the future sensitivity of the LZ experiment~\cite{Mount:2017qzi} (note that the green dotted line coincides with the black diagonal line in the right panel).
In the left panel, the DM candidate is the lighter CP-even scalar $X_1$,
while in the right panel the black diagonal line separates the region where the DM candidate is $X_1$ from the region where the DM candidate is the CP-odd scalar $a^0$, as denoted by the labels ``CP-even'' and ``CP-odd''.
Gray dashed lines are contours of the DM candidate mass $m_\mathrm{DM}$.}
\label{SD2d}
\end{figure}

In Fig.~\ref{SD2d}, we demonstrate the expected sensitivities at 95\% CL in the $m_D-m_S$ plane from current and CEPC precisions of EW oblique parameters, with solid and dot-dashed lines correspond to the assumption of $U=0$ and the assumption that all the oblique parameters are free in the global fits, respectively.
For comparison, we also plotted 90\% CL constraints from the spin-independent direct detection experiment XENON1T~\cite{Aprile:2017iyp}, as well as the expected sensitivity for a 1000-day run of the future experiment LZ~\cite{Mount:2017qzi}.
In order to give a clear demonstration of the CEPC capability, we fix some model parameters, whose values are chosen to make some regions in the $m_D-m_S$ plane evading direct detection.

In the left panel, the couplings are fixed as $\lambda_1=\lambda_{Sh}=\lambda_3=1$, $\lambda_2=-1$, and $\kappa=0.8v$, leading to a CP-even scalar DM candidate.
The direct detection constraint in the $m_S>m_D$ region is looser than in the $m_S<m_D$ region.
The reason is that the choice of coupling values here implies $\lambda_1+2\lambda_2+\lambda_3=0$. Therefore, from Eq.~\eqref{SDtrilinear} we know that the $hX_1X_1$ coupling vanishes in the limit $c_\theta \to 0$, which is reached when $m_S\gg m_D$ and $X_1$ is totally $\phi^0$.

In the right panel, $\lambda_2$ is set to $1$, while the other couplings have the same values as in the left panel.
In this case, we tend to have $m_1 < m_a$ for $m_S<m_D$ and $m_a<m_1$ for $m_S>m_D$.
Thus, the DM candidate basically is $X_1$ in the $m_S<m_D$ region, but becomes the CP-odd scalar $a^0$ in the $m_S>m_D$ region.
Since the coupling values lead to $\lambda_1-2\lambda_2+\lambda_3=0$ and hence a vanishing $ha^0a^0$ coupling, the $m_S>m_D$ region escapes from direct detection constraints.

In both panels, the expected CEPC sensitivities are almost independent of $m_S$, because the singlet $S$ does not have any gauge interaction. They can explore the parameter space up to $m_D\sim 1~\si{TeV}$. Therefore, CEPC would effectively cover the regions with $m_S>m_D$ where direct detection experiments may not adequately probe.

\subsection{Reduction to the Inert Higgs Doublet Model}

In the limit $\kappa=0$ and $m_S\to\infty$, the singlet $S$ decouples and the SDSDM model reduces to the inert Higgs doublet model~\cite{Deshpande:1977rw}.
Then the mass spectrum becomes
\begin{equation}
m_{\phi^0}^2=m_D^2+\frac{1}{2}(\lambda_1+2\lambda_2+\lambda_3)v^2,\quad
m_a^2=m_D^2+\frac{1}{2}(\lambda_1-2\lambda_2+\lambda_3)v^2,\quad
m_C^2=m_D^2+\frac{1}{2}\lambda_1v^2.
\end{equation}
The mass splitting is just controlled by $\lambda_2$ and $\lambda_3$.
When $\lambda_2<0$ $(\lambda_2>0)$, the CP-even scalar $\phi^0$ (the CP-odd scalar $a^0$) is the DM candidate, which is lighter than $\phi^\pm$ when $\lambda_3<2|\lambda_2|$.
The $h\phi^0\phi^0$ and $ha^0a^0$ couplings are given by
\begin{equation}\label{IHDtrilinear}
\mathcal{L}\supset -\frac{1}{2}(\lambda_1+2\lambda_{2}+\lambda_{3})vh(\phi^0)^2
-\frac{1}{2}(\lambda_1-2\lambda_{2}+\lambda_{3})vh(a^0)^2.
\end{equation}
The contributions to $\Pi_{IJ}(p^2)$ reduce to Eqs.~\eqref{eq:D:Pi_33}--\eqref{eq:D:Pi_11}.

\begin{figure}[!t]
\centering
\includegraphics[width=0.49\textwidth]{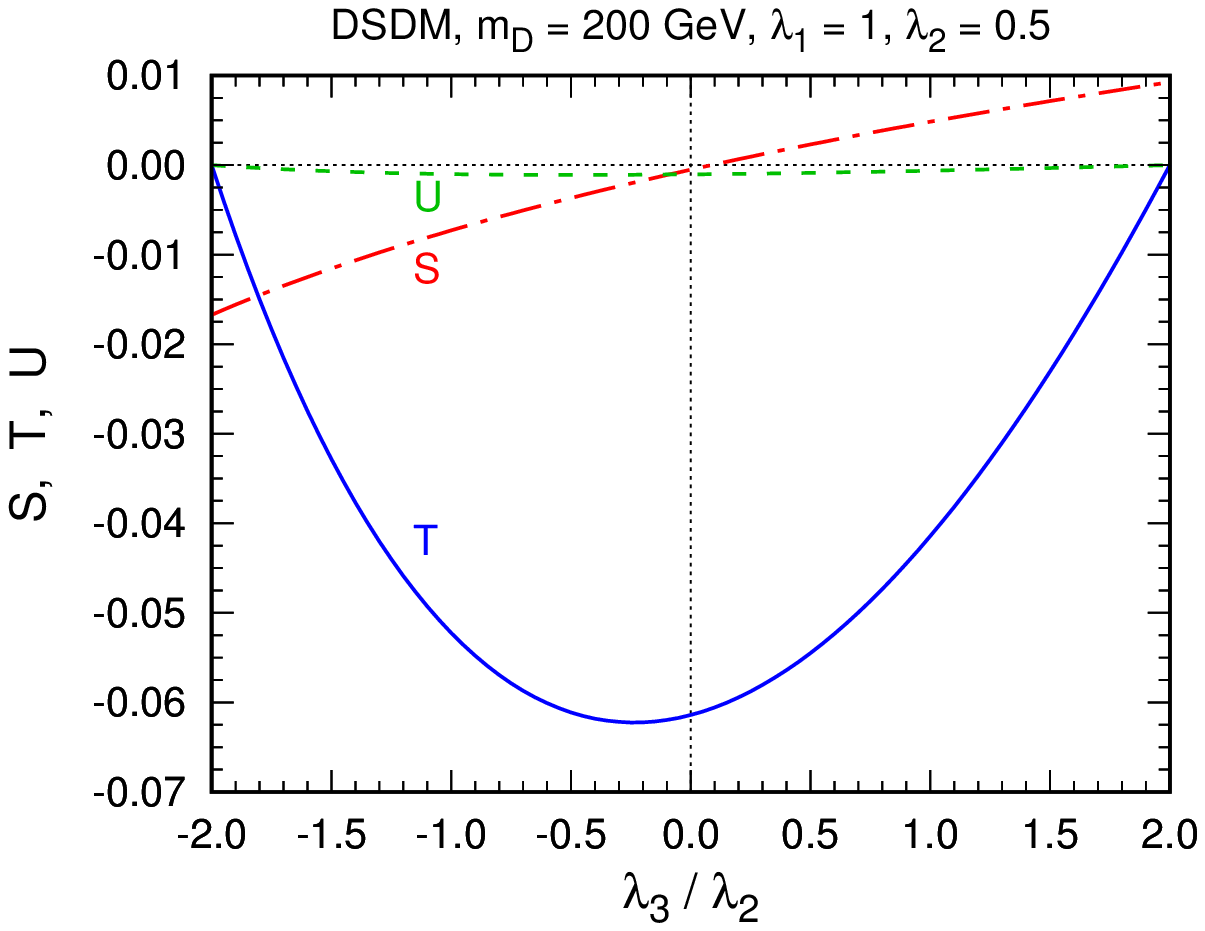}
\includegraphics[width=0.49\textwidth]{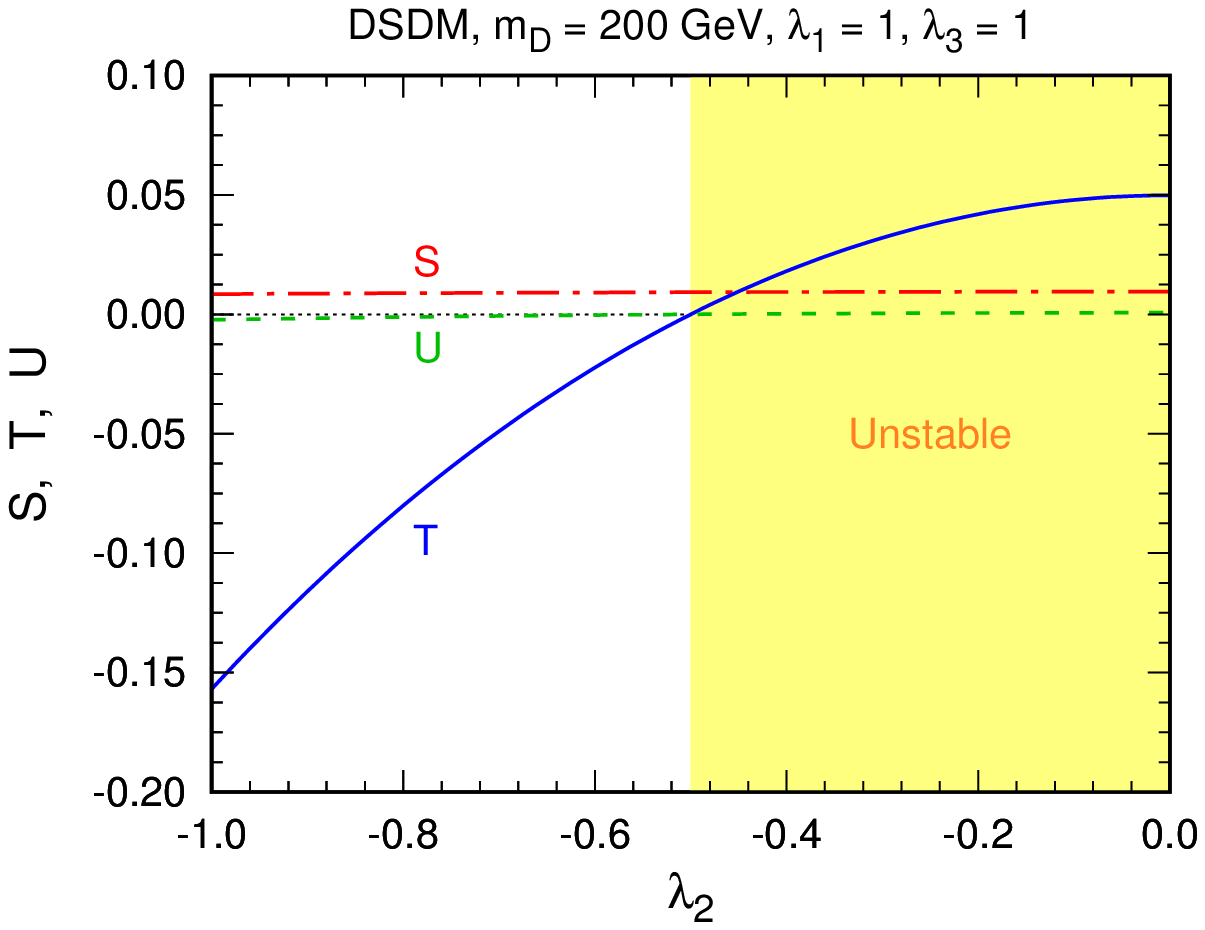}
\caption{$S$, $T$, and $U$ as functions of $\lambda_3/\lambda_2$ (left) and $\lambda_2$ (right) in the inert Higgs doublet model. In the left panel, the parameters are fixed as $m_D=200~\si{GeV}$, $\lambda_1=1$, and $\lambda_2=0.5$,
while in the right panel, they are fixed as $m_D=200~\si{GeV}$ and $\lambda_1=\lambda_3=1$.
The yellow color denotes the region where the DM candidate is unstable because it is heavier than the charged scalar $\phi^\pm$.}
\label{DSTU}
\end{figure}

Fig.~\ref{DSTU} demonstrates the behavior of EW oblique parameters.
As shown in the left panel, $T$ and $U$ vanish in the custodial symmetry limits $\lambda_3/\lambda_2=\pm 2$, while $S$ positively correlates to $\lambda_3$.
In the right panel, $S$ is basically independent of $\lambda_2$, but $T$ positively correlates to $\lambda_2$. $T$ and $U$ reach zero at $\lambda_2=\lambda_3/2$ due to the custodial symmetry.
Note that when $-\lambda_3/2<\lambda_2<0$,  $\phi^\pm$ is the lightest particle in the dark sector and there is no stable DM candidate.

\begin{figure}[!t]
\centering
\includegraphics[width=0.49\textwidth]{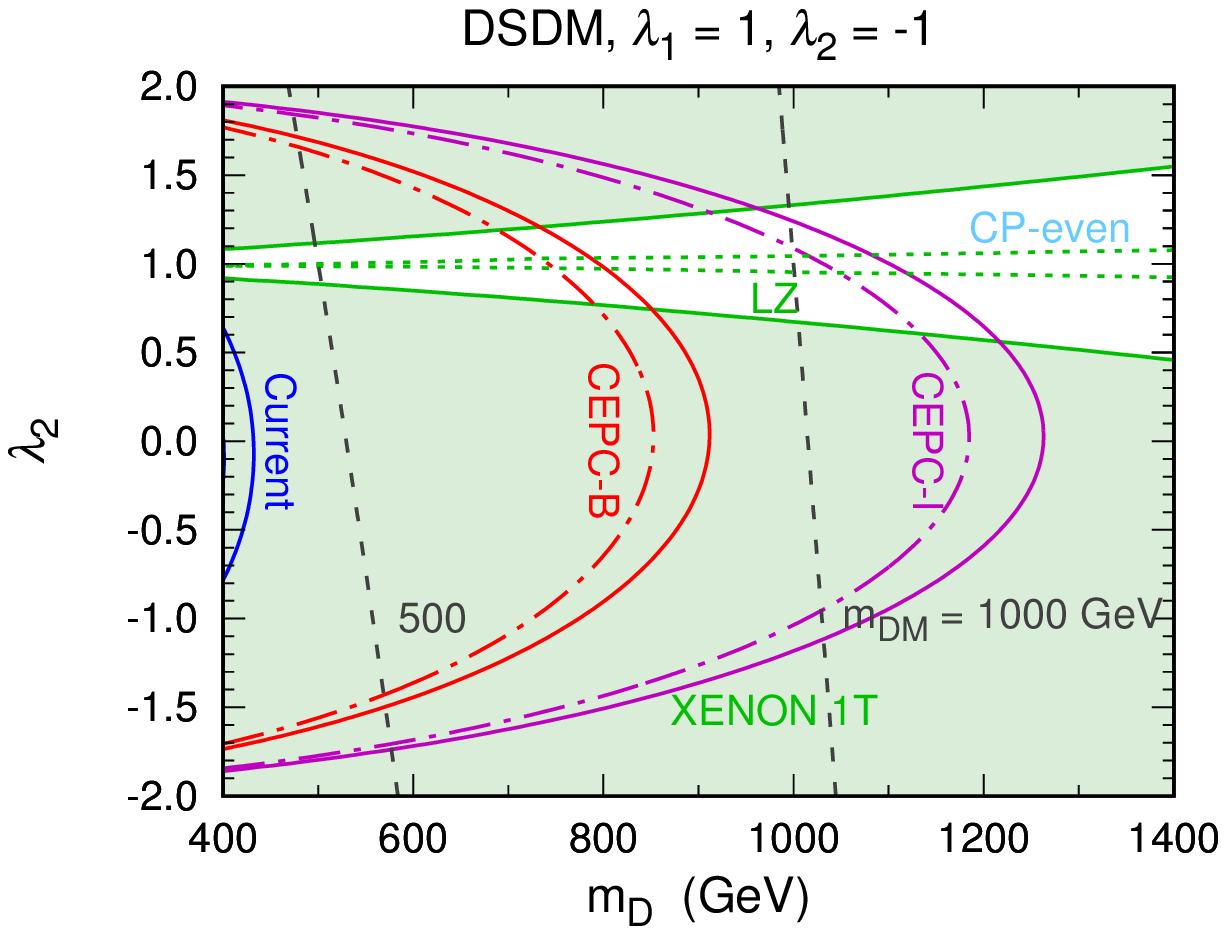}
\includegraphics[width=0.49\textwidth]{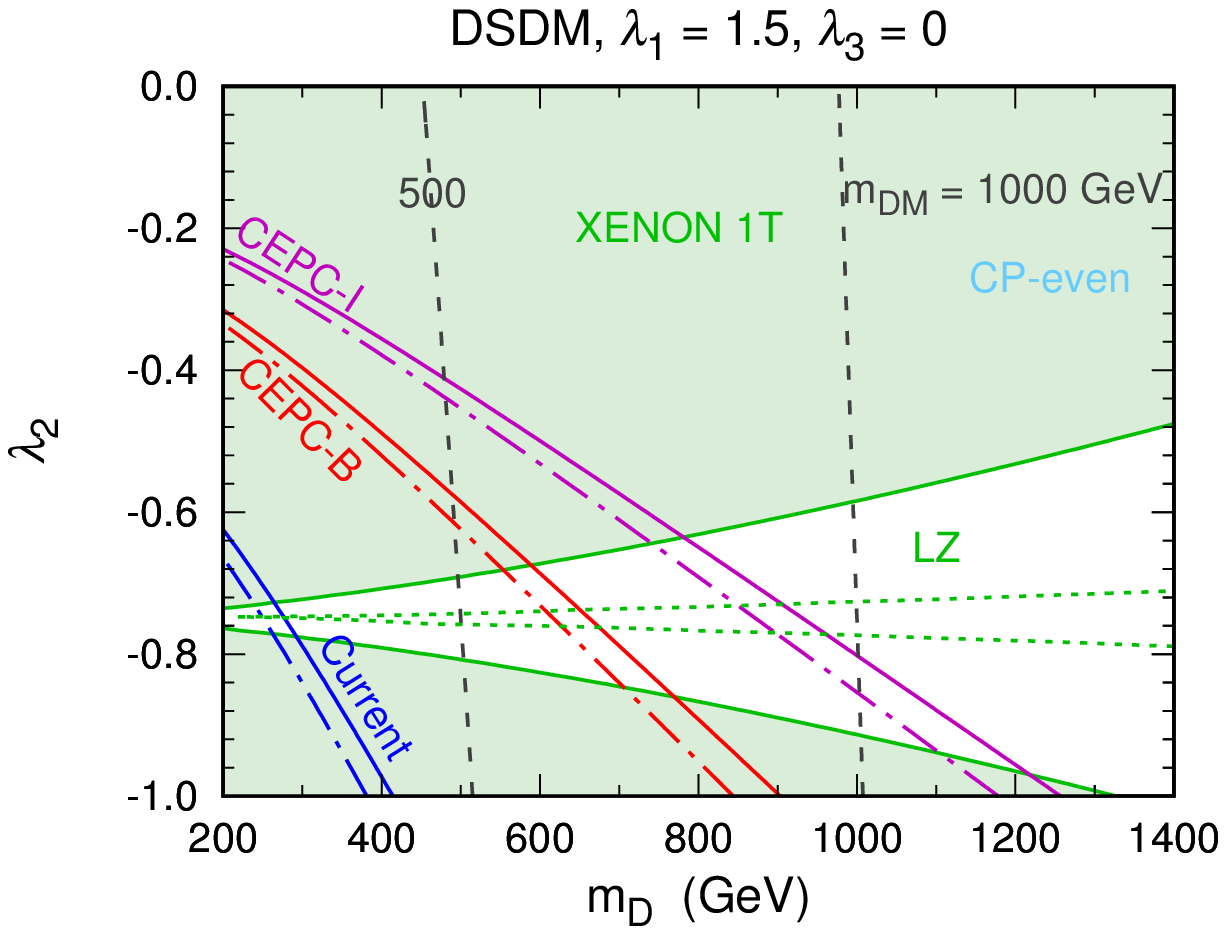}
\caption{Expected 95\% CL sensitivities  from CEPC precisions of EW oblique parameters for the inert Higgs doublet model.
In the left (right) panel, the result is presented in the $m_D-\lambda_3$ ($m_D-\lambda_2$) plane with fixed couplings $\lambda_1=1$ and $\lambda_2=-1$ ($\lambda_1=1.5$ and $\lambda_3=0$).
In both panels, the DM candidate is the CP-even scalar $\phi^0$.
The meanings of colors, labels, and line types are the same as in Fig.~\ref{SD2d}.}
\label{D2d}
\end{figure}

In Fig.~\ref{D2d}, we present the expected CEPC sensitivities and current direct detection constraints when $\phi^0$ is the DM candidate.
Direct detection experiments have excluded a large portion of parameter space, and the LZ experiment would explore further.
Nonetheless, there are some remaining regions where $\lambda_1+2\lambda_{2}+\lambda_{3}$ is small, leading to a weak $h\phi^0\phi^0$ coupling.
Future CEPC experiments can give a complementary test on such regions, probing up to a $\sim\si{TeV}$ mass scale.

\section{Singlet-Triplet Scalar Dark Matter}
\label{sec:STSDM}

\subsection{Fields and Interactions}
In the STSDM model, a real singlet scalar $S$ and a complex triplet scalar $\Delta$ with zero hypercharge are introduced:
\begin{equation}
S\in (\mathbf{1},0), \quad
\Delta=\begin{pmatrix}\Delta^+\\ \Delta^0\\ \Delta^- \end{pmatrix}=\begin{pmatrix}\Delta^{11}\\ \sqrt{2}\Delta^{12}\\-\Delta^{22} \end{pmatrix}\in(\mathbf{3},0),\quad
\Delta^0=\frac{\phi^0+ia^0}{\sqrt{2}}.
\end{equation}
Here we have provided the dictionary between the vector notation ($\Delta^+$, $\Delta^0$, $\Delta^-$) and the tensor notation ($\Delta^{ij}$)  for the triplet. See Appendix~\ref{app:conv_tensor} for further information about these notations.
Note that $(\Delta^\pm)^*$ and $\Delta^\mp$ are different scalars, since we assume $\Delta$ is complex.

The relevant Lagrangian is
\begin{equation}
\mathcal{L}\supset\frac{1}{2}(\partial_\mu S)^2+(D_\mu\Delta)^\dag D^\mu\Delta-V(S,\Delta),
\end{equation}
where $D_\mu=\partial_\mu-igW_\mu^a\tau_\mathrm{(3)}^a$ with $\tau_\mathrm{(3)}^a$ denoting the generators for $\mathbf{3}$.
The gauge interaction terms for $\Delta$ can be found in Eq.~\eqref{eq:ST:gauge}.
The scalar potential $V(S,\Delta)$ is given by
\begin{eqnarray}
V(S,\Delta)&=&m_\Delta^2|\Delta|^2+\frac{1}{2}m_S^2S^2
+\frac{\lambda_{Sh}}{2}S^2|H|^2
+\lambda_0|H|^2|\Delta|^2
+\lambda_1H^i\Delta^\dag_{ij}\Delta^{jk}H^\dag_k\nonumber\\
&&+\lambda_2H^\dag_{j'}\Delta^\dag_{ij}\Delta^{ik}H^{k'}\epsilon_{kk'}\epsilon^{j'j}
+\Big(\lambda_3 H^\dag_i\Delta^{ij}\Delta^{j'k}H^{k'}\epsilon_{jj'}\epsilon_{kk'}
+\lambda'_3|H|^2\Delta^{ij}\Delta^{i'j'}\epsilon_{ii'}\epsilon_{jj'}
\nonumber\\
&&+\lambda_4 SH^\dag_i\Delta^{ij}H^{j'}\epsilon_{jj'}
+\mathrm{h.c.}\Big)
+(\text{irrelevant terms}).
\end{eqnarray}

In the unitary gauge, we have
\begin{equation}
\lambda_3 H^\dag_i\Delta^{ij}\Delta^{j'k}H^{k'}\epsilon_{jj'}\epsilon_{kk'}
\to  - \frac{{{\lambda _3}}}{2}{(v + h)^2}\left[ {{\Delta ^ - }{\Delta ^ + } + \frac{1}{2}{{({\Delta ^0})}^2}} \right]
\end{equation}
and
\begin{equation}
\lambda'_3|H|^2\Delta^{ij}\Delta^{i'j'}\epsilon_{ii'}\epsilon_{jj'}
\to  - \frac{{{\lambda'_3}}}{2}{(v + h)^2}[2{\Delta ^ - }{\Delta ^ + } + {({\Delta ^0})^2}].
\end{equation}
As these two terms just differ by a constant factor, the effect of $\lambda'_3$ can be absorbed into $\lambda_3$.
Moreover, by defining $\lambda_\pm\equiv\lambda_1\pm\lambda_2$, the $\lambda_1$ and $\lambda_2$ terms can be expressed as
\begin{eqnarray}
&&{\lambda _1}H_i^\dag \Delta _j^i({\Delta ^\dag })_k^j{H^k} + {\lambda _2}H_i^\dag ({\Delta ^\dag })_j^i\Delta _k^j{H^k}\nonumber\\
&&\qquad\to \frac{{{\lambda _ + }}}{4}{(v + h)^2}(|{\Delta ^ + }{|^2} + |{\Delta ^ - }{|^2} + |{\Delta ^0}{|^2}) + \frac{{{\lambda _ - }}}{4}{(v + h)^2}(|{\Delta ^ - }{|^2} - |{\Delta ^ + }{|^2}).\qquad
\end{eqnarray}
Since
\begin{equation}
{\lambda _0}|H{|^2}|\Delta {|^2} \to \frac{{{\lambda _0}}}{2}{(v + h)^2}(|{\Delta ^ + }{|^2} + |{\Delta ^ - }{|^2} + |{\Delta ^0}{|^2}),
\end{equation}
$\lambda_0$ can be further absorbed into $\lambda_+$.
Therefore, we can safely neglect $\lambda'_3$ and $\lambda_0$ without affecting the discussion below.
Thus, the mass terms in the dark sector after EW symmetry breaking are given by
\begin{equation}
\mathcal{L}_\mathrm{mass}=-\frac{1}{2}(S\quad\phi^0)M_0^2\begin{pmatrix}S\\ \phi^0\end{pmatrix}
-\frac{1}{2}m_a^2(a^0)^2
-\Big((\Delta^+)^\ast\quad\Delta^-\Big)M_C^2\begin{pmatrix}\Delta^+\\(\Delta^-)^\ast\end{pmatrix},
\end{equation}
with
\begin{eqnarray}
M_0^2&=&\begin{pmatrix}m_S^2+\dfrac{1}{2}\lambda_{Sh}v^2&-\dfrac{1}{2}\lambda_4v^2\\[.8em]
-\dfrac{1}{2}\lambda_4v^2&m_\Delta^2+\dfrac{1}{4}(\lambda_+-2\lambda_3)v^2\end{pmatrix},\quad
m_a^2=m_\Delta^2+\frac{1}{4}(\lambda_+ +2\lambda_3)v^2,\qquad\\[.4em]
M_C^2&=&\begin{pmatrix}m_\Delta^2+\dfrac{1}{4}(\lambda_+-\lambda_-)v^2&-\dfrac{1}{2}\lambda_3v^2\\-\dfrac{1}{2}\lambda_3v^2&m_\Delta^2+\dfrac{1}{4}(\lambda_+ +\lambda_-)v^2\end{pmatrix}.
\end{eqnarray}
Charged and neutral mass eigenstates are obtained through the following relations with rotation matrices $U$ and $V$:
\begin{eqnarray}
\begin{pmatrix}\Delta^+\\ (\Delta^-)^\ast\end{pmatrix}&=&U\begin{pmatrix}\Delta_1^+\\ \Delta_2^+\end{pmatrix},\quad
U^\mathrm{T}M_C^2U=\begin{pmatrix}m_1^2&0\\0&m_2^2\end{pmatrix},\quad
U=\begin{pmatrix}c_\theta&-s_\theta\\s_\theta&c_\theta\end{pmatrix},\\
\begin{pmatrix}S\\ \phi^0\end{pmatrix}&=&V\begin{pmatrix}X_1\\ X_2\end{pmatrix},\quad
V^\mathrm{T}M_0^2V=\begin{pmatrix}\mu_1^2&0\\0&\mu_2^2\end{pmatrix},\quad
V=\begin{pmatrix}c_\alpha&-s_\alpha\\s_\alpha&c_\alpha\end{pmatrix}.
\end{eqnarray}
The masses of the charged scalars $\Delta_1^\pm$ and $\Delta_2^\pm$ are
\begin{equation}
m_{1,2}^2=m_\Delta^2+\frac{v^2}{4}\Big(\lambda_+\mp\sqrt{\lambda_-^2+4\lambda_3^2}\Big),
\end{equation}
while the masses of the CP-even neutral scalars $X_1$ and $X_2$ are determined by
\begin{equation}
\mu_{1,2}^2=\frac{1}{2}\Big\{(M_0^2)_{11}+(M_0^2)_{22}\mp\sqrt{[(M_0^2)_{11}-(M_0^2)_{22}]^2+4(M_0^2)_{12}^2}\Big\}.
\end{equation}

There are three situations for the DM candidate in this model.
\begin{itemize}
\item $\mu_1<m_a,m_1$: the DM candidate is the lighter CP-even scalar $X_1$.
\item $m_a<\mu_1,m_1$: the DM candidate is the CP-odd scalar $a^0$.
\item $m_1<\mu_1,m_a$: no stable DM candidate.
\end{itemize}
DM-nucleon scatterings can be induced by the $hX_1X_2$ and $ha^0a^0$ couplings, whose forms in the Lagrangian are
\begin{equation}
\mathcal{L}\supset -\left[\frac{\lambda_{Sh}}{2}c_\alpha^2-\lambda_4s_\alpha c_\alpha+\frac{1}{4}(\lambda_+ -2\lambda_{3})s_\alpha^2\right]vhX_1^2
-\frac{1}{4}(\lambda_+ +2\lambda_{3})vh(a^0)^2.
\end{equation}

We calculate the one-loop corrections to EW gauge boson vacuum polarizations from dark sector scalars in the STSDM model, and the results are given by Eqs.~\eqref{eq:ST:Pi_3Q} and \eqref{eq:ST:Pi_11}.
Note that the condition $\Pi_{3Q}(p^2)=\Pi_{33}(p^2)$ holds, resulting in $S=0$.
This is reasonable, since neither the singlet nor the triplet carries any hypercharge.
If the model also respects the custodial symmetry, $T$ and $U$ would vanish and this model could not be explored via EW oblique parameters.

\subsection{Custodial Symmetry}

Below we identify the condition for the custodial symmetry.
As the singlet $S$ and the triplet $\Delta$ have zero hypercharge, both of them are trivial under the $\mathrm{SU}(2)_\mathrm{R}$ global transformation.
Defining an $\mathrm{SU}(2)_\mathrm{R}$ vector $\mathcal{H}^I$ with $\mathcal{H}^1 = \tilde{H}$ and $\mathcal{H}^2 = H$, we can see that
\begin{equation}
\mathcal{H}^\dag_{I,i} \mathcal{H}^{I,j}
=H^\ast_{j'}H^{i'}\epsilon_{i'i}\epsilon^{jj'}+H^\ast_iH^j
\end{equation}
is an $\mathrm{SU}(2)_\mathrm{R}$ invariant.
By using the identity~\eqref{identity}, the generic scalar potential that respects the custodial symmetry can be expressed as
\begin{eqnarray}
V_\mathrm{cust}&=&\mathcal{H}^\dag_{I,i} \mathcal{H}^{I,j}\delta^i_m\delta^n_j(\lambda_aS\Delta^{mk}\epsilon_{nk}+\lambda_b\Delta^{mk}\Delta^\ast_{kn}+\lambda_c\Delta^{mk}\Delta^{k'n'}\epsilon_{kk'}\epsilon_{nn'}+\mathrm{h.c.})\nonumber\\
&=&(H^\ast_{j'}H^{i'}\epsilon_{i'i}\epsilon^{jj'}+H^\ast_iH^j)\left[2(\tau^a)^i_j(\tau^a)^n_m+\frac{1}{2}\delta^i_j\delta^n_m\right]\nonumber\\
&&\times(\lambda_aS\Delta^{mk}\epsilon_{nk}+\lambda_b\Delta^{mk}\Delta^\ast_{kn}+\lambda_c\Delta^{mk}\Delta^{k'n'}\epsilon_{kk'}\epsilon_{nn'}+\mathrm{h.c.})\nonumber\\
&=&|H|^2(\lambda_aS\Delta^{mk}\epsilon_{mk}+\lambda_b|\Delta|^2+\lambda_c\Delta^{mk}\Delta^{k'n'}\epsilon_{kk'}\epsilon_{mn'}+\mathrm{h.c.}).
\end{eqnarray}
The last equality comes from the fact that the identity $\epsilon_{i'i}(\tau^a)^i_j\epsilon^{jj'}=-(\tau^{a})^{j'}_{i'}$
leads to $(H^\ast_{j'}H^{i'}\epsilon_{i'i}\epsilon^{jj'}+H^\ast_iH^j)(\tau^a)^i_j=0$.
In the bracket of the last line, the first term vanishes because the symmetric tensor $\Delta^{mk}$ is contracted with the antisymmetric tensor $\epsilon_{mk}$.
This means that the custodial symmetry implies $\lambda_4=0$.
The second term implies a condition that $\lambda_1=\lambda_2$, and the third term is just the $\lambda_3$ and $\lambda_{3'}$ terms.
In conclusion, the condition for the custodial symmetry is
\begin{equation}
\lambda_4=\lambda_-=0.
\end{equation}

In the limit $\lambda_4=0$ and $m_S\to\infty$,
the singlet $S$ decouples and the mass spectrum becomes
\begin{equation}
\mu_1^2 = m_\Delta ^2 + \frac{1}{4}({\lambda _ + } - 2{\lambda _3}){v^2},~~
m_a^2 = m_\Delta ^2 + \frac{1}{4}({\lambda _ + } + 2{\lambda _3})v,~~
m_{1,2}^2=m_\Delta^2+\frac{v^2}{4}\Big(\lambda_+\mp\sqrt{\lambda_-^2+4\lambda_3^2}\Big).
\end{equation}
Consequently, if $\lambda_-\neq 0$, we will have $m_1^2<\min(\mu_1^2,m_a^2)$ and there will be no stable DM candidate.
If $\lambda_- = 0$, the DM candidate will degenerate in mass with $\Delta_1^\pm$. However, $\lambda_- = 0$ also leads to the custodial symmetry and $S=T=U=0$.
This is not the case we are interested in.

\subsection{Expected Sensitivity}

\begin{figure}[!t]
\centering
\includegraphics[width=0.49\textwidth]{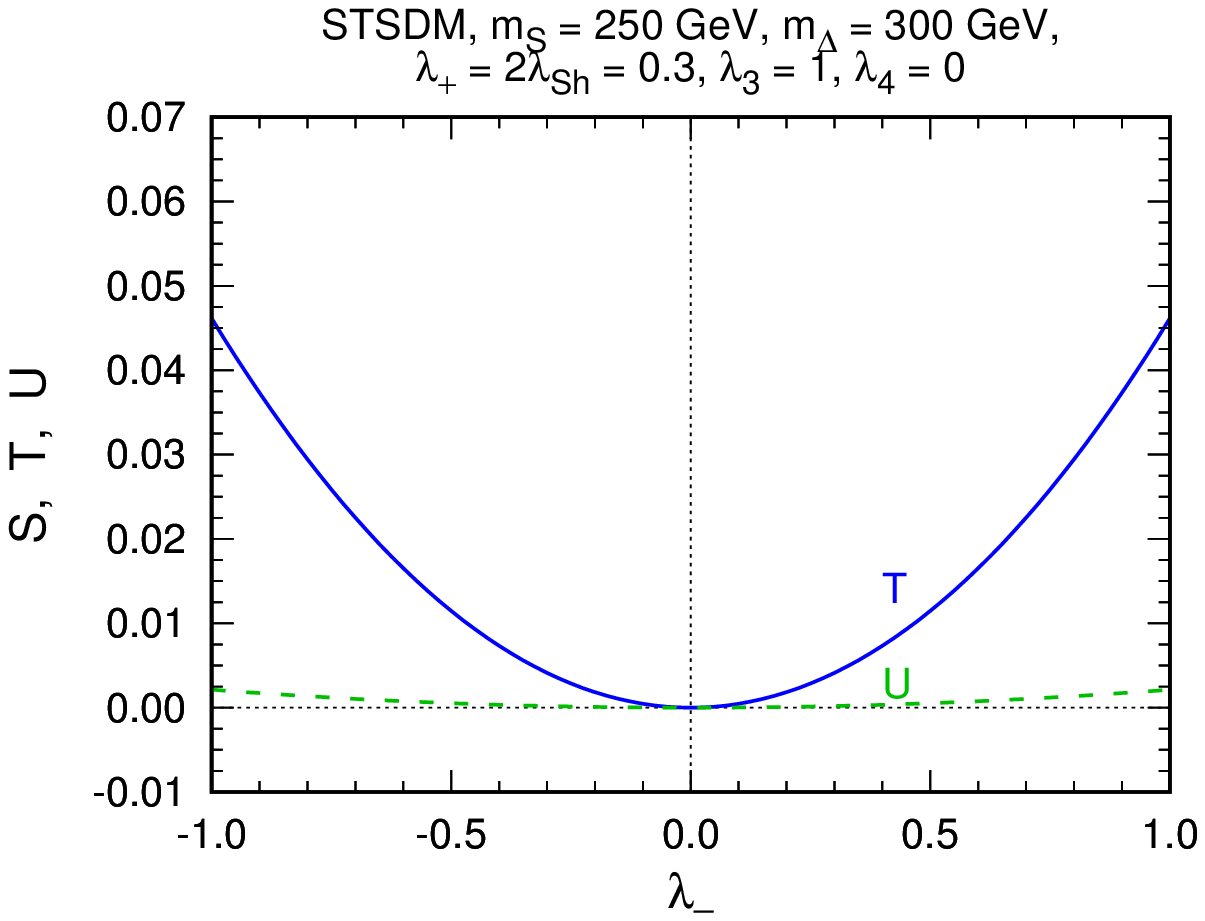}
\includegraphics[width=0.49\textwidth]{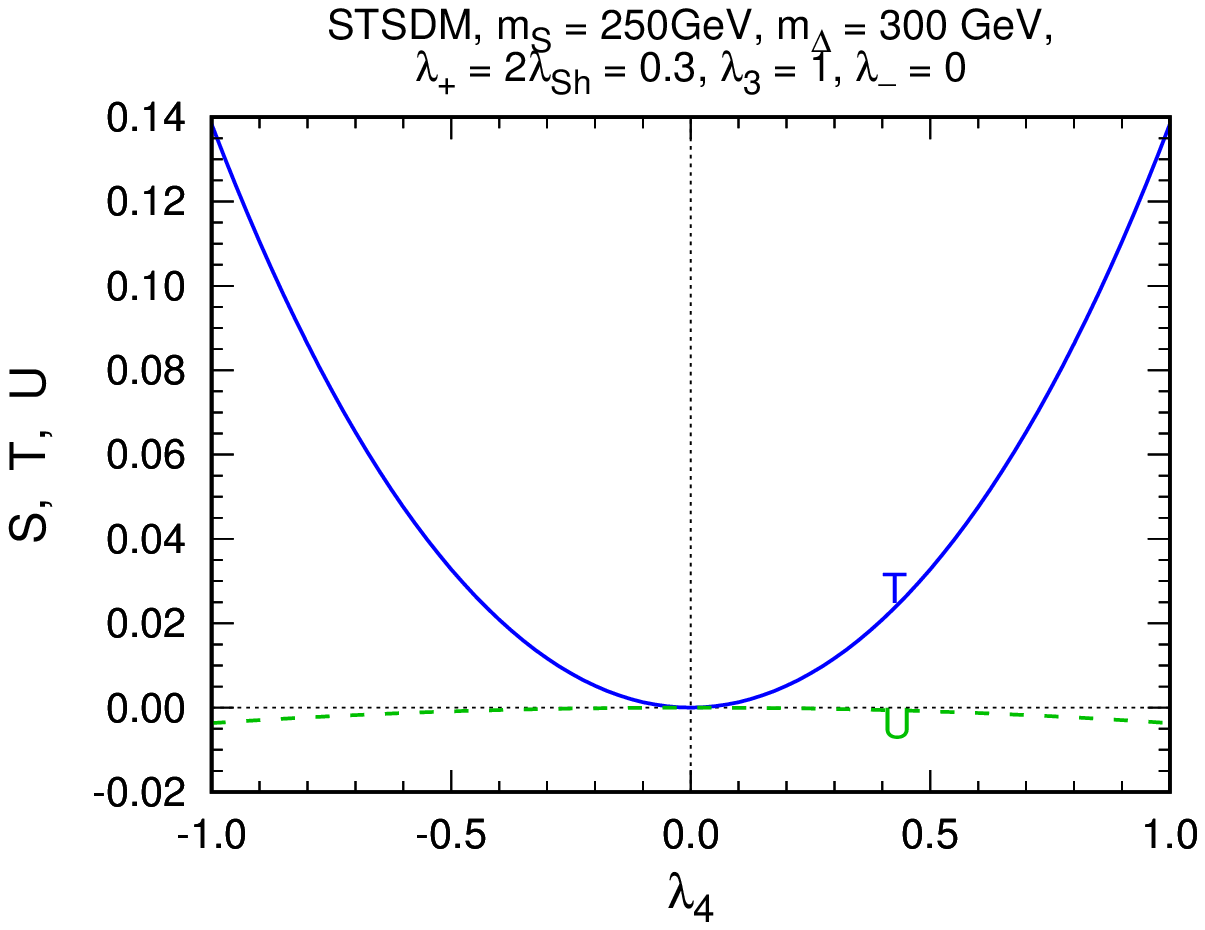}
\caption{$T$ and $U$ as functions of $\lambda_-$ (left) and $\lambda_4$ (right) in the STSDM model. In the left (right) panel, the parameters are fixed as $m_S=250~\si{GeV}$, $m_\Delta=300~\si{GeV}$, $\lambda_+ = 2\lambda_{Sh}=0.3$, $\lambda_3=1$, and $\lambda_4=0$ ($\lambda_1=0$).}
\label{STSTU}
\end{figure}

In Fig.~\ref{STSTU}, we demonstrate the behaviors of $T$ and $U$ depending on $\lambda_-$ or $\lambda_4$, with other parameters fixed.
Obviously, $T$ and $U$ vanish at $\lambda_4=\lambda_-=0$, due to the the custodial symmetry. As $S=0$ in this model, the dominant deviation from the SM comes from the $T$ parameter, which increases as $\lambda_4$ or $\lambda_-$ increases. EW precision measurements would be more sensitive to the regions with large $\lambda_4$ and $\lambda_-$.

\begin{figure}[!t]
\centering
\includegraphics[width=0.49\textwidth]{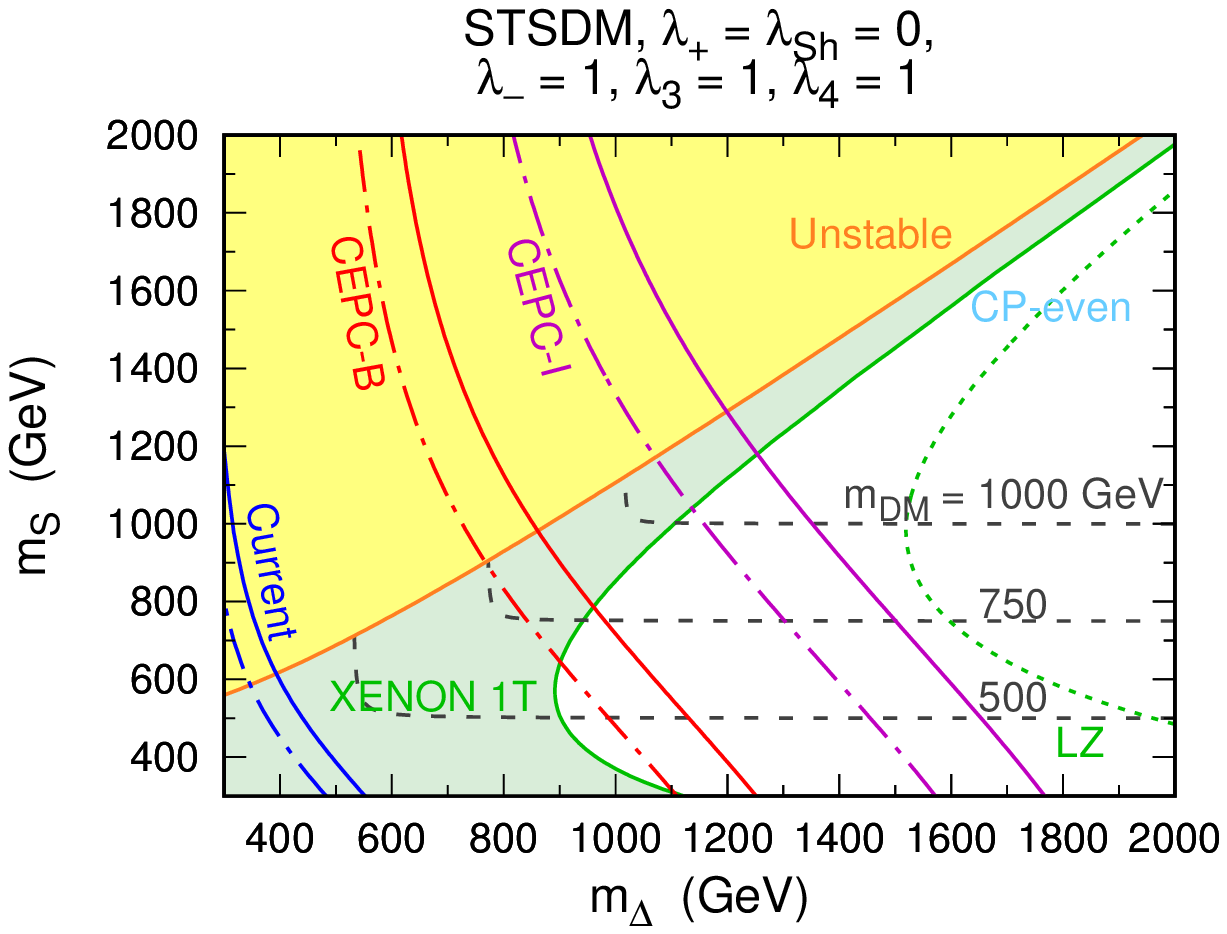}
\includegraphics[width=0.49\textwidth]{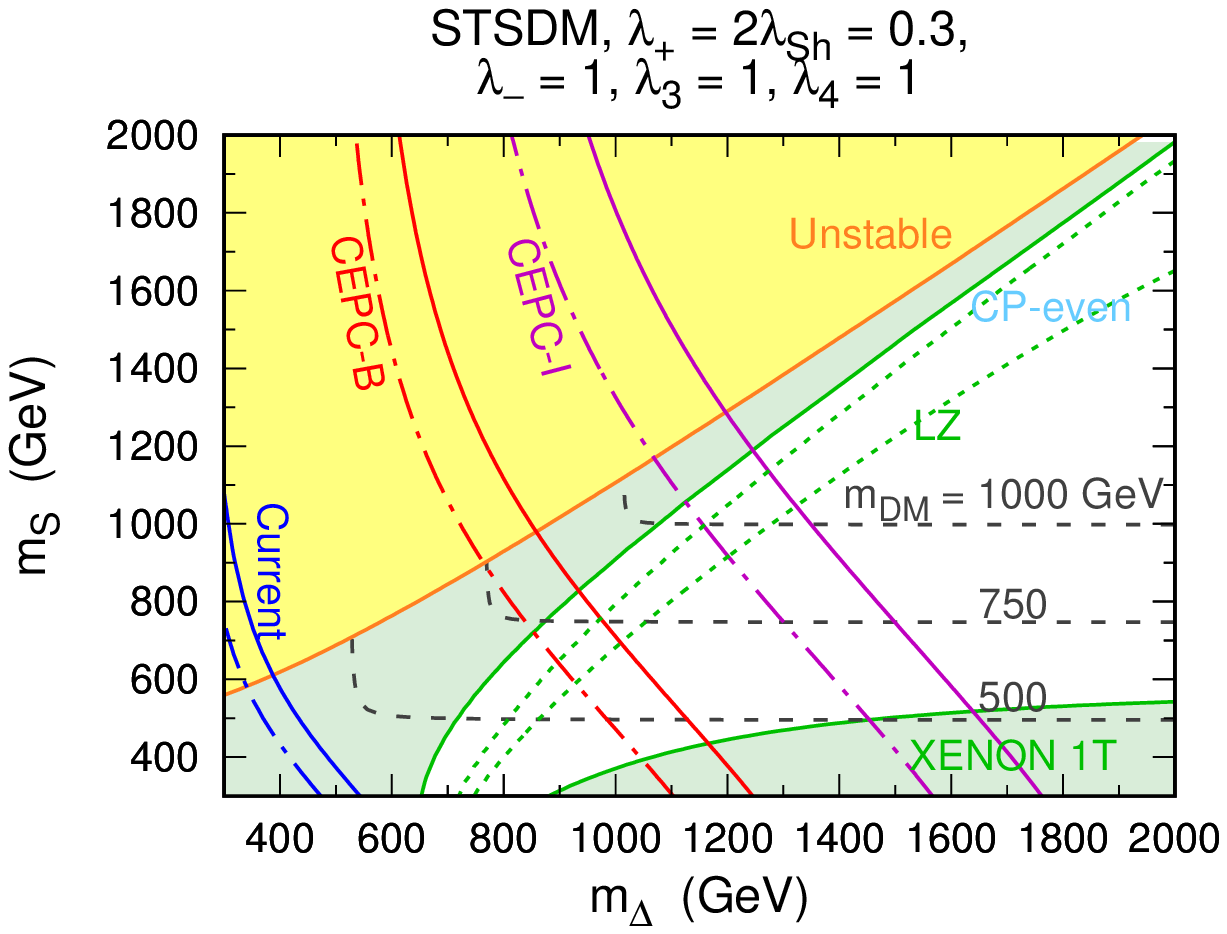}
\caption{Expected 95\% CL sensitivities from CEPC precisions of EW oblique parameters in the $m_\Delta-m_S$ plane for the STSDM model.
In the left (right) panel, the parameters are fixed as $\lambda_+ = 2\lambda_{Sh}=0~(0.3)$, $\lambda_-=\lambda_3=\lambda_4=1$.
The meanings of colors, labels and line types are the same as in Fig.~\ref{SD2d}, except that the solid and dot-dashed lines for the sensitivities from oblique parameters correspond to the assumptions of $S=U=0$ and $S=0$ in the global fits, respectively. There is no stable DM candidate in the yellow regions, while the DM candidate in the remaining regions is the CP-even scalar $X_1$.}
\label{ST2d}
\end{figure}

Fig.~\ref{ST2d} shows the expected sensitivities in the $m_\Delta-m_S$ plane from CEPC precisions of EW oblique parameters, as well as the bounds from current direct detection experiments.
Since the STSDM model predicts $S=0$, we estimate these expected sensitivities based on the precisions under the assumptions of $S=U=0$ and for $S=0$ in the global EW fits.
For the parameters we choose, the $m_S > m_\Delta$ half plane is unlikely to provide a stable DM candidate, while the CP-even scalar $X_1$ appears as a viable DM candidate in the other half plane.

In the left panel with $\lambda_+=\lambda_{Sh}=0$, current direct detection can hardly exclude the region with the DM particle mass $m_\mathrm{DM}\gtrsim 300~\si{GeV}$.
CEPC EW measurements can test the model up to $m_\mathrm{DM}\sim 1~\si{TeV}$, but the CEPC sensitivity would not be better than LZ.
In the right panel with $\lambda_+=2\lambda_{Sh}=0.3$, direct detection experiments can probe the right bottom region where $m_\Delta>1~\si{TeV}$ and $m_S<600~\si{GeV}$ corresponding to $m_\mathrm{DM}<500$~\si{GeV}. The CEPC sensitivity for this case is quite similar to the left panel. Nonetheless, it will probe a narrow region that would not be covered by the LZ experiment.

\section{Quadruplet Scalar Dark Matter}
\label{sec:QSDM}

\subsection{Fields and Interactions}
In QSDM model, we consider a dark sector consisting of a complex quadruplet scalar $X$ with $Y=1/2$:
\begin{equation}
X=\begin{pmatrix}X^{++}\\ X^+\\ X^0 \\X^- \end{pmatrix}=\begin{pmatrix}X^{111}\\ \sqrt{3}X^{112}\\ \sqrt{3}X^{122}\\ X^{222}\end{pmatrix}\in(\mathbf{4},1/2),\quad
X^0=\frac{\phi^0+ia^0}{\sqrt{2}},
\end{equation}
where the dictionary between the vector notation ($X^{++}$, $X^+$, $X^0$, $X^-$) and the tensor notation ($X^{ijk}$) is explicitly given.
Note that $(X^\pm)^*\neq X^\mp$, since $X$ is complex.
We have the related Lagrangian
\begin{equation}
\mathcal{L}\supset(D_\mu X)^\dag D^\mu X-V(X)
\end{equation}
with $D_\mu=\partial_\mu-igW_\mu^a\tau_{(4)}^a-ig'B_\mu/2$ and
\begin{eqnarray}
V(X)&=&m_X^2|X|^2+\lambda_0|H|^2|X|^2
+\lambda_1H^\dag_iX^{ijk}X^\dag_{jkl}H^l
+\lambda_2H^{i'}X^{ijk}X^\dag_{jkl}H^\dag_{l'}\epsilon_{i'i}\epsilon^{ll'}\nonumber\\
&&+(\lambda_3X^{imn}X^{jm'n'}H^\dag_iH^\dag_j\epsilon_{mm'}\epsilon_{nn'}+\mathrm{h.c.})+(\text{irrelevant terms}).
\end{eqnarray}
The gauge interactions of $X$ are given by Eq.~\eqref{eq:Q:gauge}.

By defining $\lambda_\pm\equiv\lambda_1\pm\lambda_2$, one can check that $\lambda_0$ can be absorbed into $\lambda_+$.
Without loss of generality, we have the following mass terms in the unitary gauge:
\begin{equation}
\mathcal{L}_\mathrm{mass}=-\frac{1}{2}m_\phi^2(\phi^0)^2
-\frac{1}{2}m_a^2(a^0)^2
-\Big((X^+)^\ast\quad X^-\Big)M_C^2\begin{pmatrix}X^+\\(X^-)^\ast\end{pmatrix}
-m_{++}^2|X^{++}|^2,
\end{equation}
where
\begin{eqnarray}
m_\phi^2&=&m_X^2+\frac{1}{12}(3\lambda_++\lambda_--8\lambda_3)v^2,\quad
m_a^2=m_X^2+\frac{1}{12}(3\lambda_++\lambda_-+8\lambda_3)v^2,\\
M_C^2&=&\begin{pmatrix}m_X^2+\dfrac{1}{12}(3\lambda_+-\lambda_-)v^2&\dfrac{1}{\sqrt{3}}\lambda_3v^2\\[1em] \dfrac{1}{\sqrt{3}}\lambda_3v^2&m_X^2+\dfrac{1}{4}(\lambda_++\lambda_-)v^2\end{pmatrix},\\
m_{++}^2&=&m_X^2+\frac{1}{4}(\lambda_+-\lambda_-)v^2.
\end{eqnarray}
The mass eigenstates of the singly charged scalars are related to the gauge eigenstates via a rotation matrix $U$:
\begin{equation}
\begin{pmatrix}X^+\\ (X^-)^\ast\end{pmatrix}=U\begin{pmatrix}X_1^+\\ X_2^+\end{pmatrix},\quad
U^\mathrm{T}M_C^2U=\begin{pmatrix}m_1^2&\\&m_2^2\end{pmatrix},\quad
U=\begin{pmatrix}c_\theta&-s_\theta\\s_\theta&c_\theta\end{pmatrix},
\end{equation}
where the mass eigenvalues are
\begin{equation}
m_{1,2}^2=m_X^2+\frac{1}{12}\left(3\lambda_++\lambda_-\mp 2\sqrt{\lambda_-^2+12\lambda_3^2}\right)v^2.
\end{equation}

As the DM candidate should be the lightest particle in the dark sector, there are three situations.
\begin{itemize}
\item $\lambda_3>0$ and $|\lambda_-|\leq 2\lambda_3$: the DM candidate is the CP-even scalar $\phi^0$;
\item $\lambda_3<0$ and $|\lambda_-|\leq -2\lambda_3$: the DM candidate is the CP-odd scalar $a^0$;
\item $|\lambda_-|>2|\lambda_3|$: there is no stable DM candidate.
\end{itemize}
DM direct detection experiments are sensitive to the $h\phi^0\phi^0$ and $ha^0a^0$ couplings
\begin{equation}\label{eq:QSDM:hXX}
\mathcal{L}\supset -\frac{1}{12}(3\lambda_++\lambda_--8\lambda_3)vh(\phi^0)^2
-\frac{1}{12}(3\lambda_++\lambda_-+8\lambda_3)vh(a^0)^2.
\end{equation}


The dark sector contributions to the vacuum polarizations of EW gauge bosons can be found in Eqs.~\eqref{eq:Q:Pi_33}--\eqref{eq:Q:Pi_11}, which are used to calculate the predictions of $S$, $T$, and $U$.

\subsection{Custodial Symmetry}

In order to find the condition for the custodial symmetry, we define two $\mathrm{SU}(2)_\mathrm{R}$ vector $\mathcal{H}^I$ and $\mathcal{X}^I$ as
\begin{equation}
\mathcal{H}^{1,i} = \tilde{H}^i=\epsilon^{ii'}H^{\ast}_{i'},~~
\mathcal{H}^{2,i} = H^i,~~
\mathcal{X}^{1,ijk}=\tilde{X}^{ijk}=\epsilon^{ii'}\epsilon^{jj'}\epsilon^{kk'}X^\ast_{i'j'k'},~~
\mathcal{X}^{2,ijk}=X^{ijk}.
\end{equation}
They satisfy
\begin{equation}
\epsilon_{ii'}\epsilon_{II'}\mathcal{H}^{I',i'}=-\mathcal{H}^\dag_{I,i},\quad
\epsilon_{ii'}\epsilon_{jj'}\epsilon_{kk'}\epsilon_{II'}\mathcal{X}^{I',i'j'k'}=-\mathcal{X}^\dag_{I,ijk}.
\end{equation}
The generic custodial symmetric potential can be written down as
\begin{eqnarray}\label{VofPhiZ}
V_\mathrm{cust}&=&\lambda_a\mathcal{H}^\dag_{I,i}\mathcal{H}^{I,i}\mathcal{X}^\dag_{J,jkl}\mathcal{X}^{J,jkl}
+\lambda'_a\mathcal{H}^\dag_{I,i}\mathcal{X}^\dag_{J,i'jk}\mathcal{X}^{J,jkl}\mathcal{H}^{I,l'}\epsilon^{ii'}\epsilon_{ll'}
+\lambda_b\mathcal{H}^\dag_{I,i}\mathcal{X}^{I,ijk}\mathcal{X}^\dag_{J,jkl}\mathcal{H}^{J,l}\nonumber\\
&&+\lambda_c\mathcal{H}^\dag_{I,i}\mathcal{X}^{J,ijk}\mathcal{X}^\dag_{I',jkl}\mathcal{H}^{J',l}\epsilon^{II'}\epsilon_{JJ'}\nonumber\\
&=&4\lambda_a|H|^2|X|^2+(2\lambda_b+2\lambda'_a)H^\dag_iX^{ijk}X^\dag_{jkl}H^l+(2\lambda_c+2\lambda'_a)H^{i'}X^{ijk}X^\dag_{jkl}H^\dag_{l'}\epsilon_{i'i}\epsilon^{ll'}\nonumber\\
&&+(\lambda_b-\lambda_c)(X^{imn}X^{jm'n'}H^\dag_iH^\dag_j\epsilon_{mm'}\epsilon_{nn'}+\mathrm{h.c.}).
\end{eqnarray}
Thus, we can identify that $\lambda_0=4\lambda_a$, $\lambda_1=2\lambda_b+2\lambda'_a$, $\lambda_2=2\lambda_c+2\lambda'_a$, and $\lambda_3=(\lambda_1-\lambda_2)/2=\lambda_-/2$.
Therefore, $\lambda_-=2\lambda_3$ is a condition for the custodial symmetry.

Another condition for the custodial symmetry is $\lambda_-=-2\lambda_3$.
One can check this by instead defining $\mathcal{X}^I$ with $\mathcal{X}^{1,ijk}=-\epsilon^{ii'}\epsilon^{jj'}\epsilon^{kk'}X^\ast_{i'j'k'}$.
In conclusion, when the condition
\begin{equation}
\lambda_-=\pm2\lambda_3
\end{equation}
is satisfied, the custodial symmetry is respected, resulting in vanishing $T$ and $U$.

On the other hand, if $\lambda_-=\lambda_3=0$, all the components of $X$ will be degenerate in mass:
\begin{equation}
{\mathcal{L}_{{\mathrm{mass}}}} =  - \left( {m_X^2 + \frac{1}{4}{\lambda _ + }{v^2}} \right)\left[ {|{X^{ +  + }}{|^2} + |{X^ + }{|^2} + |{X^ - }{|^2} + \frac{1}{2}{{({\phi ^0})}^2} + \frac{1}{2}{{({a^0})}^2}} \right].
\end{equation}
In this case, the nonzero Higgs VEV would not essentially break the gauge symmetry of $X$, leading to $S=T=U=0$.

\subsection{Expected Sensitivity}

\begin{figure}[!t]
\centering
\includegraphics[width=0.49\textwidth]{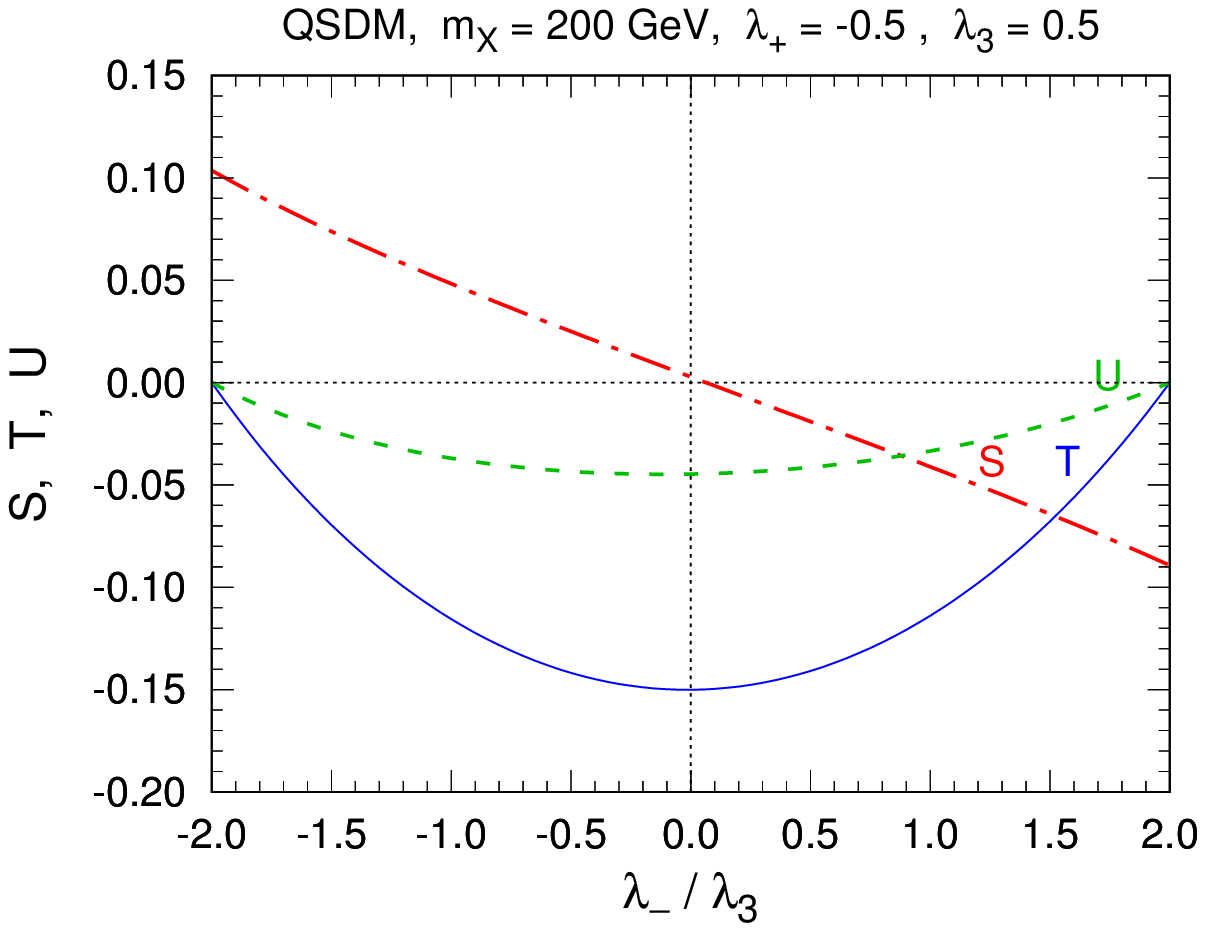}
\includegraphics[width=0.49\textwidth]{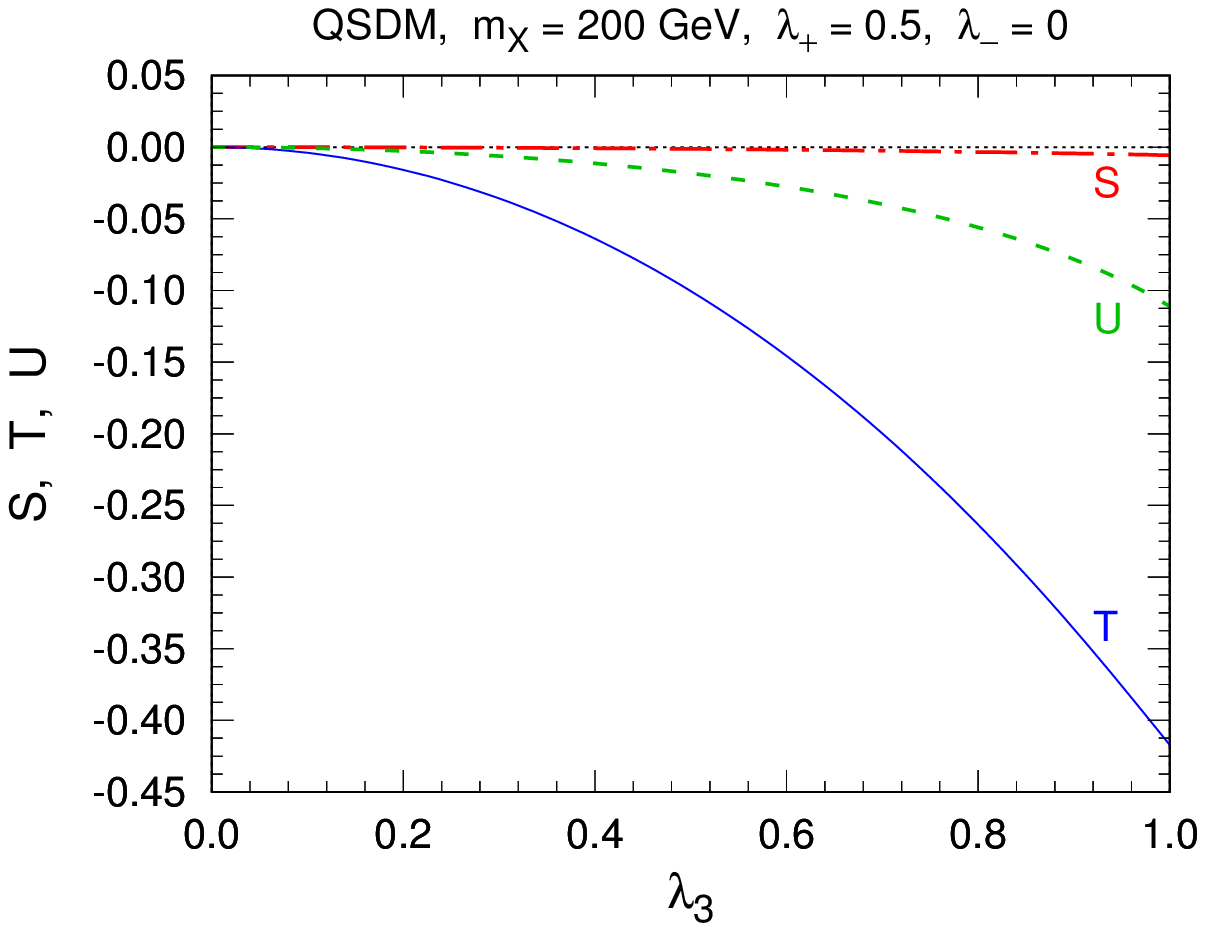}
\caption{$S$, $T$, and $U$ as functions of $\lambda_-/\lambda_3$ (left) and $\lambda_3$ (right) in the SDSDM model.
In the left (right) panel, the parameters are fixed as $m_X=200~\si{GeV}$, $\lambda_+=-0.5$, and $\lambda_3=0.5$
($m_X=200~\si{GeV}$, $\lambda_+=0.5$, and $\lambda_-=0$).}
\label{QSTU}
\end{figure}

The left and right panels of Fig.~\ref{QSTU} show the EW oblique parameters as functions of $\lambda_-/\lambda_3$ and $\lambda_3$ with other parameters fixed, respectively.
As we expected, in the left panel $T$ and $U$ vanish at the points $\lambda_-/\lambda_3=\pm 2$ that respect the custodial symmetry.
In the right panel with $\lambda_-=0$, $S$, $T$, and $U$ all vanish at the point $\lambda_3=0$.
It should be noted that $U$ can be as large as a quarter of $T$, and thus the $U=0$ assumption may not be a good approximation.

\begin{figure}[!t]
\centering
\includegraphics[width=0.49\textwidth]{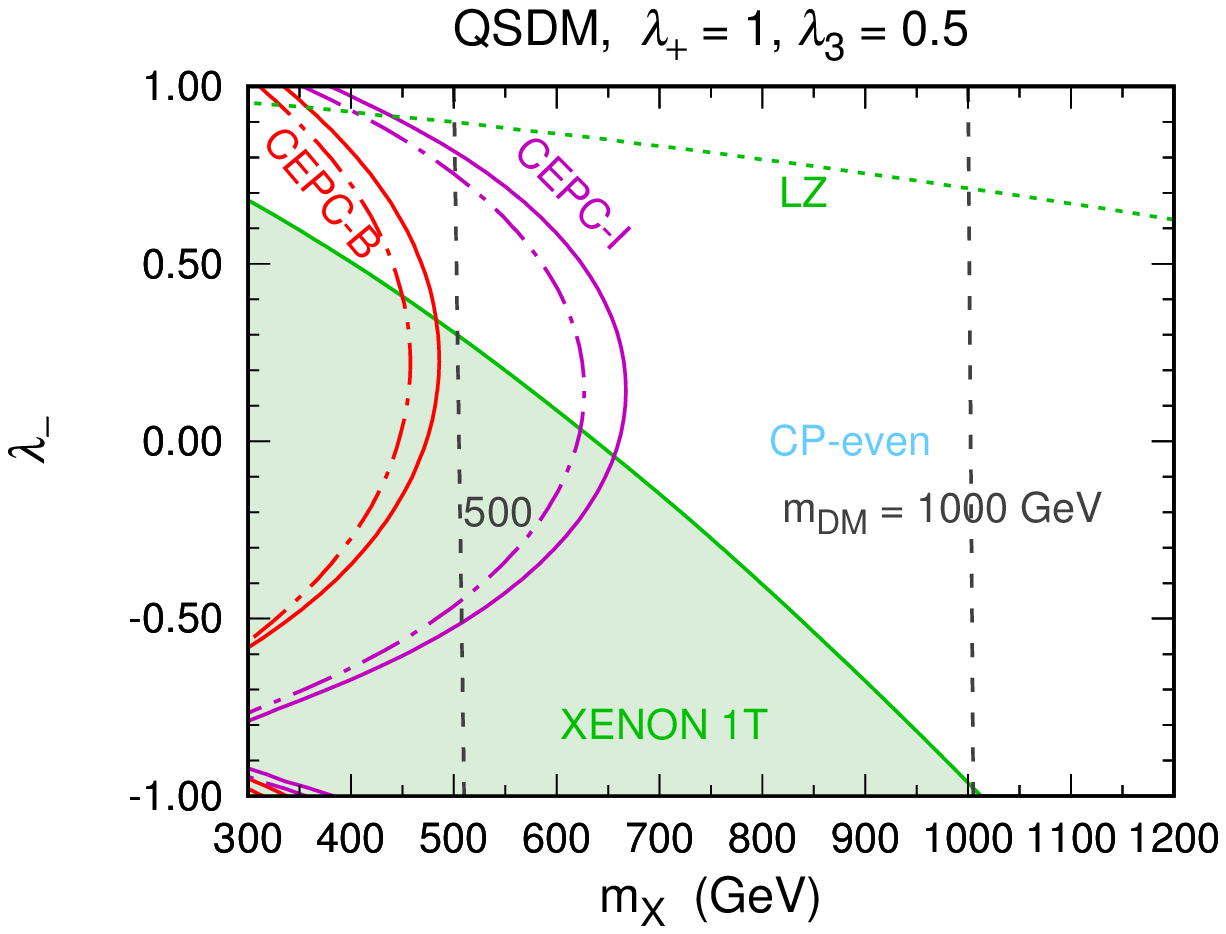}
\includegraphics[width=0.49\textwidth]{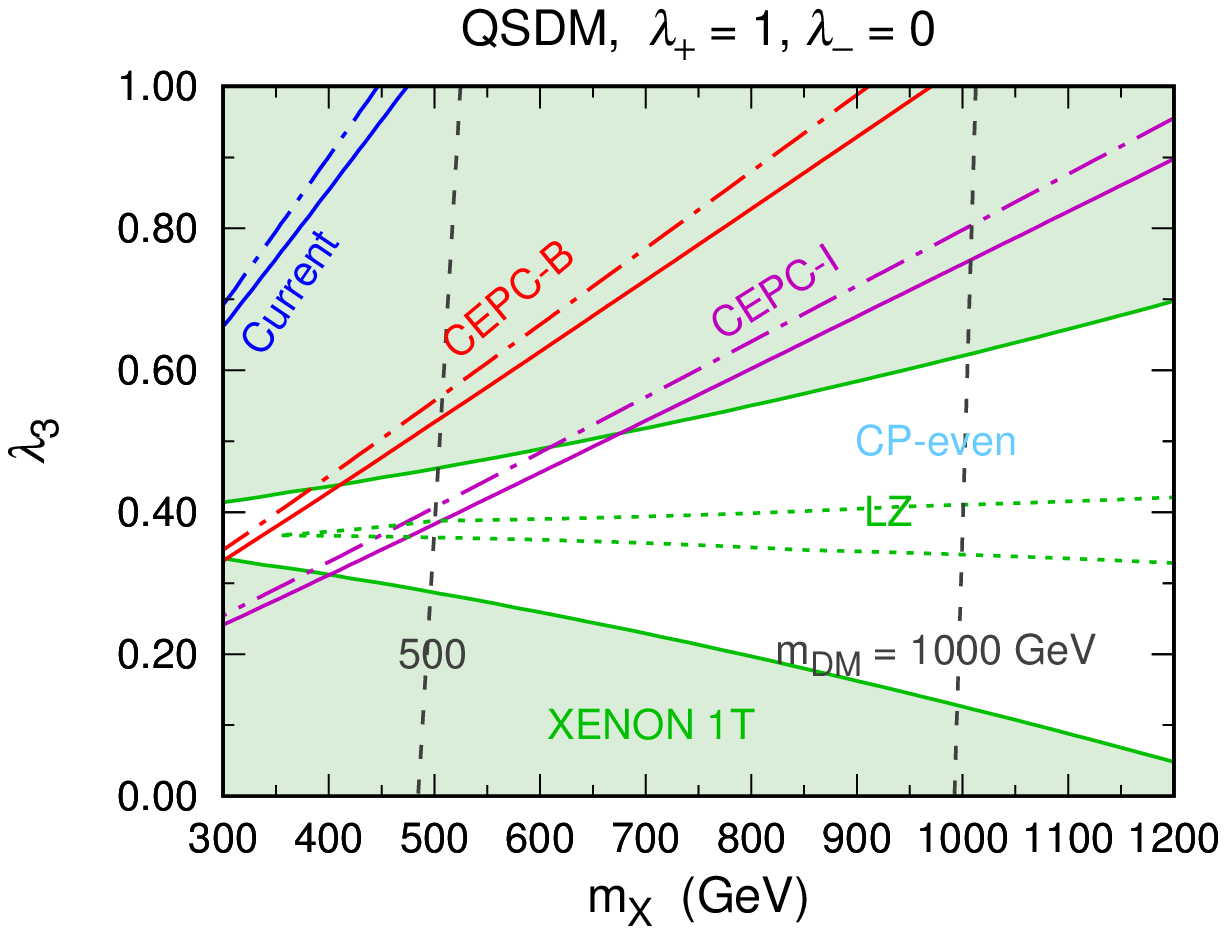}
\caption{Expected 95\% CL sensitivities from CEPC precisions of EW oblique parameters for the QSDM model.
In the left (right) panel, the result is presented in the $m_X-\lambda_-$ ($m_X-\lambda_3$) plane with fixed couplings $\lambda_+=1$ and $\lambda_3=0.5$ ($\lambda_-=0$).
The DM candidate is the CP-even scalar $\phi^0$ in both panels.
The meanings of colors, labels, and line types are the same as in Fig.~\ref{SD2d}.}
\label{Q2d}
\end{figure}

In the left (right) panel of Fig.~\ref{Q2d}, we demonstrate the expected sensitivities on the QSDM model from the CEPC measurement of EW oblique parameters in the $m_X-\lambda_-$ ($m_X-\lambda_3$) plane with $\lambda_+=1$ and $\lambda_3=0.5$ ($\lambda_-=0$) fixed.
Current constraints from direct detection experiments are also plotted.
In both panels, $\phi^0$ is the DM candidate, since we just show the results for $\lambda_3>0$.
If we choose $\lambda_3<0$ to make $a^0$ become the DM candidate, the resulting plots would be quite similar to the $\lambda_3>0$ case, and thus we do not show them here.

In the left panel, the direct detection bound is stringent in the $\lambda_-<0$ region, but it becomes much weaker than the CEPC sensitivity in the $\lambda_-<0$ region.
The reason is quite simple.
Eq.~\eqref{eq:QSDM:hXX} shows that the $h\phi^0\phi^0$ coupling is proportional to $(3\lambda_+ + \lambda_- - 8\lambda_3)$.
As we fix $\lambda_+=1$ and $\lambda_3=0.5$, this coupling becomes weaker and weaker as $\lambda_-$ increases from $-1$, and vanishes at $\lambda_-=1$.
CEPC can reach up to $m_X\sim 650~\si{GeV}$ at $\lambda_-\sim 0.1$.
In the right panel, we fix $\lambda_+=1$ and $\lambda_-=0$, leading a vanishing $h\phi^0\phi^0$ coupling at $\lambda_3=3/8$.
Therefore, the direct detection bound becomes weak around $\lambda_3=3/8$, the expected CEPC sensitivity can be stronger in the region with $m_X\lesssim 750~\si{GeV}$.
Compared with the future LZ experiment, however, CEPC just has better sensitivity in some small corners for both panels.

\section{Conclusions and Discussions}
\label{sec:concl}

Electroweak precision measurements will be improved to an unprecedented level in the future CEPC project.
In particular, a higher precision of EW oblique parameters can provide us an excellent opportunity to search for new physics with electroweak interactions.
In this work, we investigate the sensitivity from CEPC electroweak data to scalar WIMP DM models.
For comparison, current bounds and future sensitivity from DM direct detection experiments are also demonstrated.

We discuss three models, SDSDM, STSDM, and QSDM, where we introduce a dark sector consisting of one or two scalar $\mathrm{SU}(2)_\mathrm{L}$.
Their trilinear and quartic couplings to the SM Higgs doublet could induce mass splittings for their components, leading to nontrivial contributions to EW oblique parameters $S$, $T$, and $U$.
The DM candidate is the lightest mass eigenstate of neutral components.
It will be a CP-even scalar if it comes from the real parts of neutral components, or a CP-odd scalar if it comes from the imaginary part of a neutral component.
In most parameter regions, the couplings to the Higgs field generally split the mass of the CP-odd scalar and the masses of CP-even scalars, and hence the DM candidate is a real scalar that by itself cannot have vector current interactions with the $Z$ boson.
Therefore, DM-nucleus scatterings in direct detection experiments are induce by the trilinear couplings to the Higgs boson.

If the couplings between the Higgs doublet and the dark sector scalars satisfy some special conditions, there will be a custodial symmetry, resulting in vanishing $T$ and $U$.
In this case, CEPC EW data can only explore the parameter regions where $S$ is nonzero.
Some particular coupling relations could lead to a weak DM-Higgs coupling, which would be hardly probed in direct detection experiments, but could be covered by CEPC.
For some moderate values of couplings, we show that CEPC EW data can probe these models with a DM particle mass up to $\sim0.5-1~\si{TeV}$.

The above calculations for direct detection bounds and sensitivity are actually based on the assumption that the DM candidate in each model makes up all of dark matter in the Universe.If this is not the case, the direct detection results might be significantly changed, while the CEPC results would not.
If the DM candidate just constitutes a part of dark matter, the direct detection bounds on the models would be weakened.

Assuming DM was thermally produced and annihilated dominantly via EW interactions at the freeze-out epoch, the observed relic abundance would suggest that the DM candidate mass should be $0.54$, $2.0$, or $2.4~\si{TeV}$ if the DM candidate purely comes from a doublet, triplet, or quadruplet, respectively~\cite{Cirelli:2005uq}.
As the DM candidate in the SDSDM or STSDM model generally has a singlet component, such a mass prediction could be modified. The situation would be more complicated if a resonance or Sommerfeld effect on DM annihilation is important.
Nonetheless, higher masses would typically suppress annihilation and overproduce dark matter, contradicting the observation.
Lower masses would lead to a relic abundance lower than the observed value. In the latter case, unless there are extra nonthermal production mechanisms to reproduce the observed abundance, the DM candidate would just make up a portion of the whole dark matter, and the direct detection bounds would be weaker than what we present above.

\acknowledgments

This work is supported by the National Natural Science Foundation of China (NSFC)
under Grant Nos. 11375277, 11410301005, 11647606 and 11005163, the Fundamental Research
Funds for the Central Universities,
the Natural Science Foundation of Guangdong Province under Grant No. 2016A030313313,
and the Sun Yat-Sen University Science Foundation.
ZHY is supported by the Australian Research Council.

\appendix
\section{Convention for $\mathrm{SU}(2)$ tensors}
\label{app:conv_tensor}

There are two kinds of notations for expressing $\mathrm{SU}(2)$ multiplets.
In the vector notation, a multiplet is expressed as a vector whose components are arranged in the order of their eigenvalues for the third $\mathrm{SU}(2)$ generator.
In the tensor notation, a multiplet is instead represented by an $\mathrm{SU}(2)$ tensor.
Once we define a dictionary between the vector and tensor notations, any interaction term can be firstly written in the tensor notation and then be mapped into the vector notation.
In the main text, we have used the following convention for $\mathrm{SU}(2)$ tensors.

An $\mathrm{SU}(2)$ index for the basic representation $\mathbf{2}$ is written as a superscript. Several upper indices correspond to direct products of $\mathbf{2}$'s. A tensor with $(n-1)$ symmetric upper indices belongs to the irreducible representation $\mathbf{n}$. For instance,
\begin{equation}
H^i,\quad \Delta^{ij},\quad X^{ijk},
\end{equation}
which are symmetric under the permutation of indices, live in the representations $\mathbf{2}$, $\mathbf{3}$, and $\mathbf{4}$, respectively.
A lower index belongs to $\bar{\mathbf{2}}$, the complex conjugate of $\mathbf{2}$. Thus, the complex conjugation turns upper indices into lower indices:
\begin{equation}
H^\ast_i,\quad \Delta^\ast_{ij},\quad X^\ast_{ijk}.
\end{equation}

For two $\mathrm{SU}(2)$ doublet $H$ and $\Phi$ in the vector notation,  the operators $H^\dag\Phi$ and $H^\dag\tau^a\Phi\tau^a$ ($\tau^a=\sigma^a/2$) transform as a singlet and a triplet, respectively.
In the tensor notation, this can be understood via the product decomposition $\bar{\mathbf{2}}\times\mathbf{2}=\mathbf{1}+\mathbf{3}$ as
\begin{equation}
H^\ast_i\Phi^i\in\mathbf{1},\quad H^\ast_i(\tau^a)^i_j\Phi^j (\tau^a)^k_l\in\mathbf{3}.
\end{equation}
Using the identity
\begin{equation}\label{identity}
(\tau^a)^i_j(\tau^a)^k_l=\frac{1}{2}\left(\delta^i_l\delta^j_k-\frac{1}{2}\delta^i_j\delta^k_l\right),
\end{equation}
we have
\begin{equation}
H^\ast_i(\tau^a)^i_j\Phi^j(\tau^a)^k_l=\frac{1}{2}H^\ast_l\Phi^k-\frac{1}{4}H^\ast_i\Phi^i\delta^k_l.
\end{equation}
This is the traceless part in the direct product of $H^*$ and $\Phi$, and hence belongs to $\mathbf{3}$, since a traceless relation between an upper index and a lower index is equivalent to a symmetric relation between two upper indices.

All $\mathrm{SU}(2)$ representations are pseudo-real, as there exists a unitary matrix $S_{(n)}$ that relates the $\mathbf{n}$ generators $\tau^a_{(n)}$ and the $\bar{\mathbf{n}}$ generators $-[\tau_{(n)}^a]^\ast$ via a similarity transformation
\begin{equation}
S^{-1}_{(n)}\tau_{(n)}^a S_{(n)}=-[\tau_{(n)}^a]^\ast.
\end{equation}
For $\textbf{2}$, this matrix is just $i\sigma^2$, which related to the Levi-Civita symbol through $\epsilon^{ij}=i(\sigma^2)^i_j$. We can define
\begin{equation}
\tilde{H}^i=\epsilon^{ii'}H^\ast_{i'}\in\mathbf{2},\quad \tilde{X}^{ijk}=\epsilon^{ii'}\epsilon^{jj'}\epsilon^{kk'}X^\ast_{i'j'k'}\in\mathbf{4} \quad \text{with}\quad \delta^i_j=\epsilon^{ii'}\epsilon_{i'j}=\epsilon_{ii'}\epsilon^{i'j}.
\end{equation}
$\tilde{H}^i$ acts in the same way as $H^i$ under $\mathrm{SU}(2)_\mathrm{L}$, but has an opposite hypercharge.

\section{Gauge Interaction Terms and Vacuum Polarizations}
\label{app:gauge_VP}

In this appendix, we give in detail the gauge interaction terms for the EW multiplets and their contributions to the vacuum polarizations of EW gauge bosons.

In the SDSDM model, the gauge interaction of the doublet $\Phi$ can be expanded as
\begin{eqnarray}
\mathcal{L}_\mathrm{gauge}&=&\frac{g}{2}[W_\mu ^ + {\phi ^ - }i\overleftrightarrow {{\partial ^\mu }}({\phi ^0} + i{a^0}) + \mathrm{h.c.}]
+ \frac{g}{{2{c_{\mathrm{W}}}}}{Z_\mu }[i{a^0}i\overleftrightarrow {{\partial ^\mu }}{\phi ^0} + (c_{\mathrm{W}}^2 - s_{\mathrm{W}}^2){\phi ^ - }i\overleftrightarrow {{\partial ^\mu }}{\phi ^ + }]
\nonumber\\
&& + e{A_\mu }{\phi ^ - }i\overleftrightarrow {{\partial ^\mu }}{\phi ^ + }
+ \frac{{{g^2}}}{4}W_\mu ^ + {W^{ - \mu }}[2{\phi ^ + }{\phi ^ - } + {({\phi ^0})^2} + {({a^0})^2}] + {e^2}{A_\mu }{A^\mu }{\phi ^ + }{\phi ^ - } \nonumber\\
&& + \frac{{{g^2}}}{{4c_{\mathrm{W}}^2}}{Z_\mu }{Z^\mu }\left[ {{{(c_{\mathrm{W}}^2 - s_{\mathrm{W}}^2)}^2}{\phi ^ + }{\phi ^ - } + \frac{1}{2}{{({\phi ^0})}^2} + \frac{1}{2}{{({a^0})}^2}} \right]
+ \frac{{eg}}{{{c_{\mathrm{W}}}}}(c_{\mathrm{W}}^2 - s_{\mathrm{W}}^2){A_\mu }{Z^\mu }{\phi ^ + }{\phi ^ - }
\nonumber\\
&&  + \left[ {\frac{{eg}}{2}W_\mu ^ + {A^\mu }{\phi ^ - }({\phi ^0} + i{a^0}) - \frac{{{g^2}s_{\mathrm{W}}^2}}{{2{c_{\mathrm{W}}}}}W_\mu ^ + {Z^\mu }{\phi ^ - }({\phi ^0} + i{a^0}) + \mathrm{h.c.}} \right],
\label{eq:SD:gauge}
\end{eqnarray}
where $\phi^-\equiv(\phi^+)^*$ and ${\varphi _1}\overleftrightarrow {{\partial ^\mu }}{\varphi _2} \equiv {\varphi _1}{\partial ^\mu }{\varphi _2} - ({\partial ^\mu }{\varphi _1}){\varphi _2}$.

The contributions to the vacuum polarizations $\Pi_{IJ}$ from the dark sector scalars are given by
\begin{eqnarray}
\Pi_{33}(p^2)&=&\frac{1}{16\pi^2}\Big\{B_{00}(p^2,m_C^2,m_C^2)+s_\theta^2B_{00}(p^2,m_a^2,m_1^2)+c_\theta^2B_{00}(p^2,m_a^2,m_2^2)\nonumber\\
&&\phantom{\frac{1}{16\pi^2}}-\frac{1}{2}A_0(m_C^2)-\frac{1}{4}A_0(m_a^2)-\frac{1}{4}[s_\theta^2A_0(m_1^2)+c_\theta^2A_0(m_2^2)]\Big\},\label{eq:SD:Pi_33}\\
\Pi_{3Q}(p^2)&=&\frac{1}{16\pi^2}[2B_{00}(p^2,m_C^2,m_C^2)-A_0(m_C^2)],\\
\Pi_{11}(p^2)&=&\frac{1}{16\pi^2}\Big\{B_{00}(p^2,m_a^2,m_C^2)+s_\theta^2B_{00}(p^2,m_1^2,m_C^2)+c_\theta^2B_{00}(p^2,m_2^2,m_C^2)\nonumber\\
&&\phantom{\frac{1}{16\pi^2}}-\frac{1}{2}A_0(m_C^2)-\frac{1}{4}A_0(m_a^2)-\frac{1}{4}[s_\theta^2A_0(m_1^2)+c_\theta^2A_0(m_2^2)]\Big\}.\label{eq:SD:Pi_11}
\end{eqnarray}
Here the Passiano-Veltman scalar functions are defined as~\cite{Passarino:1978jh,Denner:1991kt}
\begin{eqnarray}
{A_0}({m^2}) &=& \frac{{{{(2\pi Q)}^{4 - d}}}}{{i{\pi ^2}}}\int {{d^d}q} \frac{1}{{{q^2} - {m^2} + i\varepsilon }},
\\
{B_0}({p^2},m_1^2,m_2^2) &=& \frac{{{{(2\pi Q)}^{4 - d}}}}{{i{\pi ^2}}}\int {{d^d}q} \frac{1}{{({q^2} - m_1^2 + i\varepsilon )[{{(q + p)}^2} - m_2^2 + i\varepsilon ]}},
\\
{g_{\mu \nu }}{B_{00}}({p^2},m_1^2,m_2^2) + {p_\mu }{p_\nu }{B_{11}}({p^2},m_1^2,m_2^2) \hspace*{-9em} &&
\nonumber\\
&=& \frac{{{{(2\pi Q)}^{4 - d}}}}{{i{\pi ^2}}}\int {{d^d}q} \frac{{{q_\mu }{q_\nu }}}{{({q^2} - m_1^2 + i\varepsilon )[{{(q + p)}^2} - m_2^2 + i\varepsilon ]}}.\qquad
\end{eqnarray}

In the limit $\kappa=0$ and $m_S\to\infty$, the SDSDM model reduces to the inert Higgs doublet model, and Eqs.~\eqref{eq:SD:Pi_33}--\eqref{eq:SD:Pi_11} become
\begin{eqnarray}
\Pi_{33}(p^2)&=&\frac{1}{16\pi^2}\Big[B_{00}(p^2,m_C^2,m_C^2)+B_{00}(p^2,m_a^2,m_{\phi^0}^2)
-\frac{1}{2}A_0(m_C^2)-\frac{1}{4}A_0(m_a^2)\nonumber\\
&&~\phantom{\frac{1}{16\pi^2}}-\frac{1}{4}A_0(m_{\phi^0}^2)\Big],\label{eq:D:Pi_33}\\
\Pi_{3Q}(p^2)&=&\frac{1}{16\pi^2}[2B_{00}(p^2,m_C^2,m_C^2)-A_0(m_C^2)],\\
\Pi_{11}(p^2)&=&\frac{1}{16\pi^2}\Big[B_{00}(p^2,m_a^2,m_C^2)+B_{00}(p^2,m_{\phi^0}^2,m_C^2)
-\frac{1}{2}A_0(m_C^2)-\frac{1}{4}A_0(m_a^2)\nonumber\\
&&~\phantom{\frac{1}{16\pi^2}}-\frac{1}{4}A_0(m_{\phi^0}^2)\Big].\label{eq:D:Pi_11}
\end{eqnarray}

In the STSDM model, the gauge interaction of the triplet $\Delta$ has the following form:
\begin{eqnarray}
\mathcal{L}_\mathrm{gauge} &=& \frac{g}{{\sqrt 2 }}[W_\mu ^ + ({\phi ^0} + i{a^0})i\overleftrightarrow {{\partial ^\mu }}{({\Delta ^ + })^*} + W_\mu ^ + ({\phi ^0} - i{a^0})i\overleftrightarrow {{\partial ^\mu }}{\Delta ^ - } + \mathrm{h.c.}]
\nonumber\\
&&  + (e{A_\mu } + g{c_{\mathrm{W}}}{Z_\mu })[{({\Delta ^ + })^*}i\overleftrightarrow {{\partial ^\mu }}{\Delta ^ + } + {\Delta ^ - }i\overleftrightarrow {{\partial ^\mu }}{({\Delta ^ - })^*}]
\nonumber\\
&&  + {g^2}W_\mu ^ + {W^{ - \mu }}[|{\Delta ^ + }{|^2} + |{\Delta ^ - }{|^2} + {({\phi ^0})^2} + {({a^0})^2}] - {g^2}[W_\mu ^ + {W^{ + \mu }}{({\Delta ^ + })^*}{\Delta ^ - } + \mathrm{h.c.}]
\nonumber\\
&&  + ({e^2}{A_\mu }{A^\mu } + {g^2}c_{\mathrm{W}}^2{Z_\mu }{Z^\mu } + 2eg{c_{\mathrm{W}}}{A_\mu }{Z^\mu })(|{\Delta ^ + }{|^2} + |{\Delta ^ - }{|^2})
\nonumber\\
&&  - \frac{g}{{\sqrt 2 }}[W_\mu ^ + {({\Delta ^ + })^*}({\phi ^0} + i{a^0}) + W_\mu ^ + {\Delta ^ - }({\phi ^0} - i{a^0}) + \mathrm{h.c.}](e{A^\mu } + g{c_{\mathrm{W}}}{Z^\mu }).
\label{eq:ST:gauge}
\end{eqnarray}
The dark sector contributions to $\Pi_{IJ}$ are derived as
\begin{eqnarray}
\Pi_{3Q}(p^2)&=&\Pi_{33}(p^2)=\frac{1}{16\pi^2}[4B_{00}(p^2,m_1^2,m_1^2)+4B_{00}(p^2,m_2^2,m_2^2)-2A_0(m_1^2)-2A_0(m_2^2)],\nonumber\\*
\label{eq:ST:Pi_3Q}\\
\Pi_{11}(p^2)&=&\frac{1}{16\pi^2}\big\{2(1+s_{2\theta})[s_\alpha^2B_{00}(p^2,m_1^2,\mu_1^2)+c_\alpha^2B_{00}(p^2,m_1^2,\mu_2^2)]\nonumber\\
&&~\phantom{\frac{1}{16\pi^2}}+2(1-s_{2\theta})[s_\alpha^2B_{00}(p^2,m_2^2,\mu_1^2)+c_\alpha^2B_{00}(p^2,m_2^2,\mu_2^2)]\nonumber\\
&&~\phantom{\frac{1}{16\pi^2}}+2(1-s_{2\theta})B_{00}(p^2,m_1^2,m_a^2)+2(1+s_{2\theta})B_{00}(p^2,m_2^2,m_a^2)\nonumber\\
&&~\phantom{\frac{1}{16\pi^2}}-[A_0(m_1^2)+A_0(m_2^2)+A_0(m_a^2)+s_\alpha^2A_0(\mu_1^2)+c_\alpha^2A_0(\mu_2^2)]\big\}.
\label{eq:ST:Pi_11}
\end{eqnarray}

In the QSDM model, the gauge interactions of the quadruplet $X$ components can be expressed as
\begin{eqnarray}
\mathcal{L}_\mathrm{gauge}&=&
g\left[ {\frac{{\sqrt 6 }}{2}W_\mu ^ + {X^ + }i\overleftrightarrow {{\partial ^\mu }}{{({X^{ +  + }})}^*} + \sqrt{2}W_\mu ^ + X^0i\overleftrightarrow {{\partial ^\mu }}{{({X^ + })}^*}
+ \frac{{\sqrt 6 }}{2}W_\mu ^ + {X^ - }i\overleftrightarrow {{\partial ^\mu }}(X^0)^* + \mathrm{h.c.}} \right]
\nonumber\\
&&  + e{A_\mu }\big[2{({X^{ +  + }})^*}i\overleftrightarrow {{\partial ^\mu }}{X^{ +  + }} + {({X^ + })^*}i\overleftrightarrow {{\partial ^\mu }}{X^ + } - {({X^ - })^*}i\overleftrightarrow {{\partial ^\mu }}{X^ - }\big]
\nonumber\\
&&  + \frac{g}{{2{c_{\mathrm{W}}}}}{Z_\mu }\big[(3c_{\mathrm{W}}^2 - s_{\mathrm{W}}^2){({X^{ +  + }})^*}i\overleftrightarrow {{\partial ^\mu }}{X^{ +  + }}
+ (c_{\mathrm{W}}^2 - s_{\mathrm{W}}^2){({X^ + })^*}i\overleftrightarrow {{\partial ^\mu }}{X^ + }
\nonumber\\
&&\qquad\qquad~~  + i{a^0}i\overleftrightarrow {{\partial ^\mu }}{\phi ^0} - (3c_{\mathrm{W}}^2 + s_{\mathrm{W}}^2){({X^ - })^*}i\overleftrightarrow {{\partial ^\mu }}{X^ - }\big]
\nonumber\\
&&  + {g^2}W_\mu ^ + {W^{ - \mu }}\left[ {\frac{3}{2}|{X^{ +  + }}{|^2} + \frac{7}{2}|{X^ + }|^2 + \frac{7}{4}{{({\phi ^0})}^2} + \frac{7}{4}{{({a^0})}^2} + \frac{3}{2}|{X^ - }|^2} \right]
\nonumber\\
&&  + {e^2}{A_\mu }{A^\mu }(4|{X^{ +  + }}|^2 + |{X^ + }|^2 + |{X^ - }|^2) \nonumber\\
&& + \frac{{eg}}{{{c_{\mathrm{W}}}}}{A_\mu }{Z^\mu }[2{(3c_{\mathrm{W}}^2 - s_{\mathrm{W}}^2)^2}|{X^{ +  + }}|^2 + {(c_{\mathrm{W}}^2 - s_{\mathrm{W}}^2)^2}|{X^ + }|^2 + {(3c_{\mathrm{W}}^2 + s_{\mathrm{W}}^2)^2}|{X^ - }|^2]
\nonumber\\
&&  + \frac{{{g^2}}}{{4c_{\mathrm{W}}^2}}{Z_\mu }{Z^\mu }\Big[ {{(3c_{\mathrm{W}}^2 - s_{\mathrm{W}}^2)}^2}|{X^{ +  + }}|^2 + {{(c_{\mathrm{W}}^2 - s_{\mathrm{W}}^2)}^2}|{X^ + }|^2
\nonumber\\
&&\qquad\qquad\qquad + \frac{1}{2}{{({\phi ^0})}^2} + \frac{1}{2}{{({a^0})}^2} + {{(3c_{\mathrm{W}}^2 + s_{\mathrm{W}}^2)}^2}|{X^ - }|^2 \Big]
\nonumber\\
&& +\, (W^{\pm\mu} W_\mu^\pm XX,~A^\mu W_\mu^\pm XX\text{, and }Z^\mu W_\mu^\pm XX\text{ terms}).
\label{eq:Q:gauge}
\end{eqnarray}
The contributions to $\Pi_{IJ}$ from the quadruplet are
\begin{eqnarray}
\Pi_{33}&=&\frac{1}{16\pi^2}\Bigg\{9B_{00}(p^2,m_{++}^2,m_{++}^2)+(1+2s_\theta^2)^2B_{00}(p^2,m_1^2,m_1^2)\nonumber\\
&&+8s_\theta^2c_\theta^2B_{00}(p^2,m_1^2,m_2^2)+(1+2c_\theta^2)^2B_{00}(p^2,m_2^2,m_2^2)+B_{00}(p^2,m_\phi^2,m_a^2)\nonumber\\
&&-\left[\frac{9}{2}A_0(m_{++}^2)+\frac{1+8s_\theta^2}{2}A_0(m_1^2)+\frac{1+8c_\theta^2}{2}A_0(m_2^2)+\frac{1}{4}A_0(m_\phi^2)+\frac{1}{4}A_0(m_a^2)\right]\Bigg\},\nonumber\\*
\label{eq:Q:Pi_33}\\
\Pi_{3Q}&=&\frac{1}{16\pi^2}\Big\{12B_{00}(p^2,m_{++}^2,m_{++}^2)+2(1+2s_\theta^2)B_{00}(p^2,m_1^2,m_1^2)\nonumber\\
&&+2(1+2c_\theta^2)B_{00}(p^2,m_2^2,m_2^2)
-\big[6A_0(m_{++}^2)+(1+8s_\theta^2)(1+2s_\theta^2)A_0(m_1^2)\nonumber\\
&&+(1+8c_\theta^2)(1+2c_\theta^2)A_0(m_2^2)\big]\Big\},\\
\Pi_{11}&=&\frac{1}{16\pi^2}\Bigg\{6c_\theta^2B_{00}(p^2,m_{++}^2,m_1^2)+6s_\theta^2B_{00}(p^2,m_{++}^2,m_2^2)\nonumber\\
&&+(4c_\theta^2+3s_\theta^2-4\sqrt{3}s_\theta c_\theta)B_{00}(p^2,m_1^2,m_\phi^2)+(4c_\theta^2+3s_\theta^2+4\sqrt{3}s_\theta c_\theta)B_{00}(p^2,m_1^2,m_a^2)\nonumber\\
&&+(4s_\theta^2+3c_\theta^2+4\sqrt{3}s_\theta c_\theta)B_{00}(p^2,m_2^2,m_\phi^2)+(4s_\theta^2+3c_\theta^2-4\sqrt{3}s_\theta c_\theta)B_{00}(p^2,m_2^2,m_a^2)\nonumber\\
&&-\left[\frac{3}{2}A_0(m_{++}^2)+\frac{3+4c_\theta^2}{2}A_0(m_1^2)+\frac{3+4s_\theta^2}{2}A_0(m_2^2)+\frac{7}{4}A_0(m_\phi^2)+\frac{7}{4}A_0(m_a^2)\right]\Bigg\}.\nonumber\\*
\label{eq:Q:Pi_11}
\end{eqnarray}

\section{Interaction between Scalar Dark Matter and Nucleons}
\label{app:DM_scat}

Consider that a real scalar DM particle $\chi$, which is either CP-even or CP-odd, has a trilinear coupling to the Higgs boson $h$:
\begin{equation}
\mathcal{L}_{\chi\chi h}=-\lambda_{\chi\chi h}vh\chi^2,
\end{equation}
where $v$ is the Higgs VEV.
In the zero momentum transfer limit, this coupling induce an effective interaction operator between $\chi$ and SM quarks
\begin{equation}
\mathcal{L}_q=\frac{1}{2}\sum_qF_q\chi^2\bar{q}q
\end{equation}
with $F_q = 2\lambda_{\chi\chi h}m_q/m_h^2$.

This operator further gives rise to the scalar interaction between $\chi$ and nucleons
\begin{equation}
\mathcal{L}_N=\frac{1}{2}\sum_{N=p,n}F_N\chi^2\bar{N}N,
\end{equation}
where
\begin{equation}
F_N=\sum_{q=u,d,s}F_qf_q^N\frac{m_N}{m_q}+\sum_{q=c,b,t}F_qf_Q^N\frac{m_N}{m_q} = \frac{2\lambda_{\chi\chi h} m_N}{m_h^2} \left(\sum_{q=u,d,s}f_q^N+3f_Q^N\right).\\
\end{equation}
Here $f_i^N$ are nucleon form factors~\cite{Ellis:2000ds}:
\begin{eqnarray}
&&f_u^p=0.020\pm0.004,~~ f_d^p=0.026\pm0.005,~~ f_u^n=0.014\pm0.003,\\
&&f_d^n=0.036\pm0.008,~~ f_s^p=f_s^n=0.118\pm0.062,~~ f_Q^N=\frac{2}{27}\left(1-\sum_{q=u,d,s}f_q^N\right).~~
\end{eqnarray}
Consequently, there would be spin-independent DM-nucleon scatterings with a cross section~\cite{Yu:2011by}
\begin{equation}
\sigma_{\chi N}=\frac{m_N^2 F_N^2}{4\pi(m_\chi+m_N)^2}.
\end{equation}

\bibliographystyle{JHEP}
\bibliography{SDM}

\end{document}